\documentclass[twocolumn,tighten]{aastex62}
\usepackage{amssymb}

\begin{document}

\title{A Submillimeter Perspective on the GOODS Fields (SUPER GOODS).~IV.
The Submillimeter Properties of X-ray Sources in the CDF-S}

\author[0000-0002-3306-1606]{A.~J.~Barger}
\affiliation{Department of Astronomy, University of Wisconsin-Madison,
475 N. Charter Street, Madison, WI 53706, USA}
\affiliation{Department of Physics and Astronomy, University of Hawaii,
2505 Correa Road, Honolulu, HI 96822, USA}
\affiliation{Institute for Astronomy, University of Hawaii, 2680 Woodlawn Drive,
Honolulu, HI 96822, USA}

\author[0000-0002-6319-1575]{L.~L.~Cowie}
\affiliation{Institute for Astronomy, University of Hawaii,
2680 Woodlawn Drive, Honolulu, HI 96822, USA}

\author{F.~E.~Bauer}
\affiliation{Instituto de Astrof\'isica and Centro de Astroingenier\'ia, Facultad de F\'isica, 
Pontificia Universidad Cat\'olica de Chile,
Casilla 306, Santiago 22, Chile}
\affiliation{Millennium Institute of Astrophysics (MAS), Nuncio Monse{\~{n}}or S{\'{o}}tero 
Sanz 100, Providencia, Santiago, Chile} 
\affiliation{Space Science Institute,
4750 Walnut Street, Suite 205, Boulder, Colorado 80301, USA} 

\author[0000-0003-3926-1411]{J.~Gonz{\'a}lez-L{\'o}pez}
\affiliation{N\'ucleo de Astronom\'ia de la Facultad de Ingenir\'ia y Ciencias, Universidad Diego Portales,
Av. Ej\'ercito Libertador 441, Santiago, Chile}
\affiliation{Instituto de Astrof\'isica and Centro de Astroingenier\'ia, Facultad de F\'isica, 
Pontificia Universidad Cat\'olica de Chile \\
Casilla 306, Santiago 22, Chile}

\begin{abstract}
The CDF-S is the deepest X-ray image available and will
remain so for the near future.
We provide a spectroscopic (64.5\%; 64\% with spectral classifications) 
and photometric redshift catalog for the full 7~Ms sample, but
much of our analysis focuses on the central (off-axis angles $<5\farcm7$) 
region, which contains a large, faint ALMA sample of 75 $>4.5\sigma$
850~$\mu$m sources.
We measure the 850~$\mu$m fluxes at the X-ray positions using 
the ALMA images, where available, or an ultradeep SCUBA-2 map.  
We find that the full X-ray sample produces $\sim10$\% of the 850~$\mu$m 
extragalactic background light. 
We separate the submillimeter detected X-ray sources into star-forming galaxies
and AGNs using a star formation rate (SFR) versus X-ray luminosity 
calibration for high SFR galaxies.
We confirm this separation using the X-ray photon indices.
We measure the X-ray fluxes at the accurate positions of the 75 ALMA 
sources and detect 70\% at $>3\sigma$
in either the 0.5--2 or 2--7~keV bands. However, many of these may produce
both their X-ray and submillimeter emission by star formation.
Indeed, we find that only 20\% of the ALMA sources have intermediate X-ray luminosities
(rest-frame 8--28~keV luminosities of $10^{42.5}$--$10^{44}$~erg~s$^{-1}$), and
none has a high X-ray luminosity ($>10^{44}$~erg~s$^{-1}$).
Conversely, after combining the CDF-S with the CDF-N, we find
extreme star formation 
(SFR$>300~{\rm M}_\odot$~yr$^{-1}$) in some intermediate
X-ray luminosity sources but not in any high X-ray luminosity sources.
We argue that the quenching of star formation in the most
luminous AGNs may be a consequence of the clearing of gas in these sources.
\end{abstract}


\section{Introduction}
\label{secintro}
In the `starburst-active galactic nucleus (AGN) connection' picture, 
star formation and accretion occur coevally over
cosmic time and are together responsible for the
growth of galaxies and their resident supermassive black holes. 
The search for evidence in support of this picture
has generated great interest in determining the amount of dusty star formation 
taking place in sources selected from deep {\em Chandra\/} X-ray samples 
(e.g., Barger et al.\ 2001, 2015; Lutz et al.\ 2010; Shao et al.\ 2010; Harrison et al.\ 2012;
Page et al.\ 2012; Rosario et al.\ 2012; Stanley et al.\ 2015, 2018; 
Scholtz et al.\ 2018; Ramasawmy et al.\ 2019).
Unfortunately, this has been a challenging undertaking.
Much of what we know about distant, dusty galaxies
comes from single-dish submillimeter surveys, whose poor resolution
and source confusion make it both difficult to identify counterparts at other wavelengths 
and to probe the submillimeter flux levels needed to obtain $>3\sigma$ 
detections of individual X-ray sources. 
For example, most of the previous work focused on 250~$\mu$m
{\em Herschel\/} data, where the beam full-width half maximum (FWHM)
size is $\sim18''$, and the one
source per 40 beams confusion noise is 19~mJy (Nguyen et al.\ 2010).
Thus, these investigations had to
rely on averaging or stacking analyses, in which submillimeter fluxes 
are either measured at the X-ray (or optical counterpart) positions and then
averaged, or are directly measured from stacked submillimeter images of the 
X-ray (or optical counterpart) sources. All of the X-ray sources in a 
given X-ray luminosity bin contribute to the mean submillimeter flux for that bin, 
whether or not they are detected individually. 

Barger et al.\ (2015) were able to go beyond stacking analyses by using
ultradeep SCUBA-2 observations of the {\em Chandra\/} Deep Fields (CDFs; 
the latest SCUBA-2 catalogs can be found in earlier papers in the SUPER
GOODS series, i.e., Cowie et al.\ 2017 and 2018, hereafter, C17 and C18),
where the beam FWHM size is $14''$ but the confusion is only 1.65~mJy at 850\,$\mu$m.
In combination with the {\em Spitzer\/} and {\em Herschel\/}
data, they performed spectral energy distribution (SED) fits
to determine the far-infrared (FIR) luminosities of individual X-ray sources in the 2~Ms
CDF-North (CDF-N; Alexander et al.\ 2003) and the then-4~Ms 
CDF-South (CDF-S; Xue et al.\ 2011; Lehmer et al.\ 2012).
They found that the mean values of the FIR luminosity distributions were dominated by a small 
number of high-luminosity galaxies. This led them to conclude that averaging or stacking
analyses overestimate the level of star formation taking place in the bulk of the X-ray sample 
and hence should be used with caution. They also found that most of the 
host galaxies of the $L_{\rm 2-8~keV}>10^{44}$~erg~s$^{-1}$ AGNs at
$z>1$ were not strong star formers, perhaps because their 
star formation was suppressed by AGN feedback.

Some recent analyses have looked at the distribution functions
of star formation rates (SFRs) and specific SFRs (sSFRs) 
for X-ray AGNs using high-resolution
ALMA images to refine the SFRs (e.g., Stanley et al.\ 2018;
Scholtz et al.\ 2018). Scholtz et al.\ (2018) argued, consistent
with many previous analyses by this group and others, that any 
difference between intermediate and high X-ray luminosity samples
(i.e., $10^{43}<L_{\rm 2-10~keV}<10^{44}$~erg~s$^{-1}$
versus $L_{\rm 2-10~keV}>10^{44}$~erg~s$^{-1}$)
at $1.5<z<3.2$ are subtle.
However, these analyses are based on a rather small
number of $850~\mu$m detections (e.g., Scholtz et al.\ 2018
have only 8 $>4\sigma$ detections in the deep central
regions of the CDF-S, as compared to 52 in the present work
split as 36 ALMA and 16 SCUBA-2) and on the shallower 
4~Ms {\em Chandra\/} data rather than the current 7~Ms image 
(Luo et al.\ 2017; hereafter, L17).
 
In the present work, we measure the $850~\mu$m fluxes
of the X-ray sources in the L17 catalog using ALMA 
(this includes all of the brighter submillimeter sources)
and SCUBA-2.
Conversely, we measure the X-ray properties of 
the ALMA sources using the {\em Chandra\/} images.
We also provide optical/NIR spectral classifications
for the L17 sources where we have spectra.

We analyze the submillimeter distribution versus X-ray
luminosity to determine how the submillimeter
sources relate to the X-ray sources. 
We will argue that clear and significant
differences emerge between the intermediate and high 
X-ray luminosity populations: 
the intermediate X-ray luminosity
sources have a submillimeter distribution that
is much more skewed to high submillimeter fluxes.
We discuss how this might suggest gas clearing.

\section{Data and Methods}
\label{secdat}

\subsection{X-ray Data}
\label{subsecxray}
\begin{figure}[ht]
\centerline{\includegraphics[width=9cm,angle=0]{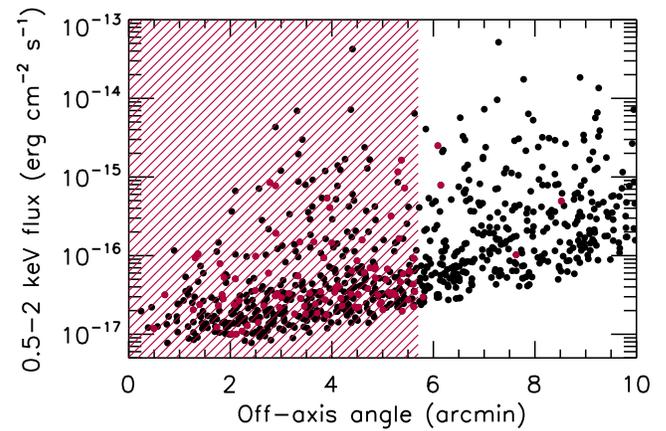}}
\caption{Observed 0.5--2~keV flux vs. {\em Chandra\/} off-axis angle for 
the X-ray sources lying within the SCUBA-2 map region where 
the 850~$\mu$m noise is less than 1.5~mJy (black circles).
The deepest SCUBA-2 observed area of 100~arcmin$^2$ 
(red shading) was also the most intensively
targeted with ALMA (non-contiguous combined ALMA area of
5.9~arcmin$^2$; see Section~\ref{subsecsubmm}).
The red circles show X-ray sources that lie in the ALMA  
images (note that some of these lie outside the red shaded region).
\label{fs_angle}
}
\end{figure}

We began with the X-ray catalog of L17 based on the 7~Ms
observations of the CDF-S. This catalog contains 1008 significant sources,
but we restrict our analysis to off-axis angles $<10'$ (hereafter referred
to as {\em the full region}) where, as we discuss below, 
a deep SCUBA-2 850~$\mu$m map exists with rms noise less than $<1.5$~mJy.
This reduces the sample to 938 X-ray sources. We hereafter refer to this as
{\em the full X-ray sample\/}. We plot observed 0.5--2~keV flux
versus off-axis angle for the full X-ray sample in Figure~\ref{fs_angle}.

In the soft (0.5--2~keV) band, the sensitivity (from L17) rises from a value
of $7\times10^{-18}$~erg~cm$^{-2}$~s$^{-1}$ on-axis, to 
$2\times10^{-17}$~erg~cm$^{-2}$~s$^{-1}$ at $5\farcm7$ 
(the edge of the region with intensive ALMA observations, hereafter
referred to as {\em the central region}; see Section~\ref{subsecsubmm}), 
to just over $10^{-16}$~erg~cm$^{-2}$~s$^{-1}$
at the $10'$ outer radius.
In the hard (2--7~keV) band, the sensitivities are about a 
factor of 5 higher. 

526 sources from the full X-ray sample lie in the central region. 
We hereafter refer to these as {\em the central region X-ray sample\/}.
We list the full X-ray sample in Table~1, ordered by {\em Chandra\/}
off-axis angle so that the most central sources appear first.

\subsection{Submillimeter Data}
\label{subsecsubmm}
The CDF-S has deep 850~$\mu$m observations made with the SCUBA-2 
camera on the JCMT, together with a large number of targeted 
ALMA band~7 observations. 
(We refer to all the submillimeter fluxes as 850~$\mu$m,
ignoring the small differences in the ALMA band~7 wavelength centers 
for different programs.)
The SCUBA-2 observations are very deep in the 100~arcmin$^2$ 
central region (red shading in Figure~\ref{fs_angle}) with 
rms noise $<0.5$~mJy throughout nearly all of the central region.
Beyond the central region,
they degrade with off-axis angle, reaching rms noise of 1.4~mJy 
at the $10'$ outer radius (see Figure~\ref{error_angle}).
In the central region, all of the $>2.25$~mJy 
850~$\mu$m SCUBA-2 sources have been observed with ALMA in band~7, 
together with a number of fainter SCUBA-2 sources. 
Details of these observations and their reductions may be found in C18. 
Note that some of the ALMA images are archival observations from 
Hodge et al.\ (2013; 4 sources), Mullaney et al.\ (2015; 8 sources),
P.I. G.~Barro (1 source), and Schreiber et al.\ (2017; 1 source).
However, we do not incorporate the four additional 
sources from the ALESS main table of Hodge et al.\ (2013) that 
correspond to X-ray sources, as they only have poor-quality cycle0 images;
they lie well outside the central region.

We restrict the area of the individual ALMA images to their
FWHM radius of $8\farcs75$. With this restriction, the ALMA
images cover a (non-contiguous) total area of 7.2~arcmin$^2$, with most
of that area (that is, 5.9~arcmin$^2$) concentrated in the central region. 
In Figure~\ref{fs_angle}, we show 0.5--2~keV flux vs. off-axis angle for  
the X-ray sources lying within the SCUBA-2 region with 
rms noise is $<1.5$~mJy (black circles).
We use red circles to denote those sources that
are also covered by the ALMA images.

As an aside, we
note that 17 of the 20 1.1~mm detected sources from the 69~arcmin$^2$
GOODS-S ALMA survey of Franco et al.\ (2018) comprise a subset of the C18
sample (with the three remaining sources considered false detections by
Franco et al.). Moreover, eight of the 12 1.2~mm detected sources from the 20~arcmin$^2$
ASAGAO ALMA survey (Ueda et al.\ 2018) are also contained in the C18 sample.
Two of their remaining four sources are not detected at the 4.5$\sigma$ level 
(S/N of 4.3 and 4.2) and have no counterparts at other wavelengths.
These sources are also not detected at the $2\sigma$ level in SCUBA-2 and
are probably false. Their other two sources (S/N of 5.2 and 4.4)
both have counterparts at other wavelengths.

Thus, the SCUBA-2 selected C18 ALMA sample recovers all reliable 
1.1--1.2~mm sources from the shallow mosaicked 
ALMA surveys of the CDF-S for a much smaller investment of ALMA time.  
Even for the deeper 4.5~arcmin$^2$ Hubble Ultra 
Deep Field ALMA survey of Dunlop et al.\ (2017), we detected
all three of their sources with 1.3~mm fluxes $>800~\mu$Jy but 
not their two fainter sources with fluxes around $300~\mu$Jy ($>5\sigma$).

\subsection{X-ray Fluxes of the ALMA Sources}
\label{subsecxrayfluxes}

\begin{figure}
\includegraphics[width=9cm,angle=0]{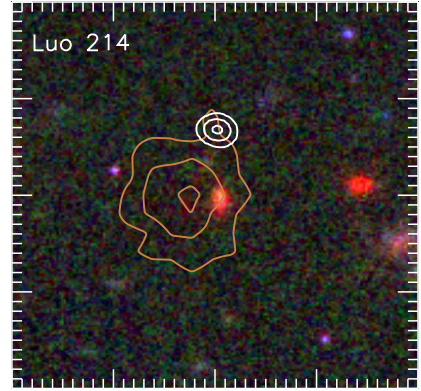}
\caption{Contour plot of L17~\#214 using the {\em Chandra\/} 
0.5--2~keV 7~Ms observations (orange contours).
The white contours show the ALMA 850~$\mu$m image of C18~\#2. 
Both sets of contours are plotted at 0.2, 0.5, and 0.9 times the 
respective maximum values. 
The contours are overlaid on the F435W (blue),
F850LP (green), and F160W (red) images from the {\em HST\/} ACS 
and CANDELS data. The image is $10''$ on a side.
\label{luo_213}
}
\end{figure}

C18 gives a catalog of 75 ALMA 
850~$\mu$m sources with extremely accurate positions (their Table~4). 
The 850~$\mu$m fluxes extend to less than 1~mJy. 
Comparing the ALMA coordinates with the L17 coordinates gives
an absolute astrometric offset of $0\farcs0$ in R.A. and $0\farcs2$
in decl. We applied this small astrometric correction to the ALMA positions.

If an ALMA source has a counterpart in the L17 catalog within $1''$,
then we take the X-ray flux from the catalog. The matching is relatively 
insensitive to the choice of matching radius, which is based on 
the $2\sigma$ positional uncertainties in the fainter {\em Chandra\/}
sources from L17. This provides X-ray fluxes for 41 of the 75 ALMA sources. 
The mean of the offsets between the ALMA and L17 positions for
these sources is $0\farcs46$.

For the remaining 34 ALMA sources, we measured the 0.5--2~keV 
or 2--7~keV fluxes in the L17 images
using the procedure outlined in Cowie et al.\ (2012) and briefly
described below. One ALMA 
source (C18~\#2) is close enough to an X-ray source 
(i.e., L17~\#214) for the X-ray flux measured at the ALMA position to be
contaminated by the X-ray emission from the neighbor (Figure~\ref{luo_213}),
so we do not use its X-ray flux.

We used a circular aperture to calculate the X-ray
fluxes, which provides a good approximation to the
point spread function (PSF)
shape at these small off-axis angles. We adopted a $1\farcs25$ aperture
radius, which offers a good compromise between including most of
the counts, maximizing the signal-to-noise (S/N), and minimizing the 
contamination from neighboring sources.
With the aperture specified, we computed the X-ray
counts~s$^{-1}$ as $C = (S-B)/t$,
where $S$ is the number of counts in the aperture,
$B(=\pi r^2 b)$ is the number of background counts expected
in the same aperture, and $t$ is the effective exposure time at the
position of this aperture. We measured the mean background 
$b$ (counts~arcsec$^{-2}$) in an 8$''$--22$''$ annulus around the source
after clipping pixels with more than 4 counts. (An extensive 
discussion of this choice may be found in Cowie et al.\ 2012.)
$C$ may be negative or positive.
We converted the counts to fluxes using a single normalization,
which we chose by comparing the aperture fluxes that we measured
for sources in the L17 catalog with the L17 fluxes.
We found good agreement with a scatter
of 23\%, which is adequate for the present work.

If an ALMA source is detected at $>3\sigma$ in either the
0.5--2~keV or 2--7~keV band, then we considered the source 
to be X-ray detected. With this procedure, 
nine more ALMA sources are X-ray detected (for a total of 50).
In Table~2, we summarize the X-ray properties of the 75 ALMA sources
(giving no X-ray fluxes for C18~\#2). 
In the 0.5--2~keV band, there are a total of 47 ALMA sources detected.
In the 2--7~keV band, there are a total of 26 ALMA
sources detected, all but 3 of which
are also detected in the 0.5--2~keV band.

\subsection{Submillimeter Fluxes of the X-ray Sources}
\label{subsecsubmmfluxes}
Where the X-ray sources
have an ALMA counterpart within the $2\sigma$ positional
uncertainties quoted by L17,
we used the best submillimeter fluxes and rms noise of C18
(Columns~8 and 9 of their Table~4).
41 X-ray sources have direct ALMA detections ($>4.5\sigma$) in
the ALMA images.

\begin{figure}[ht]
\centerline{\includegraphics[width=9cm,angle=0]{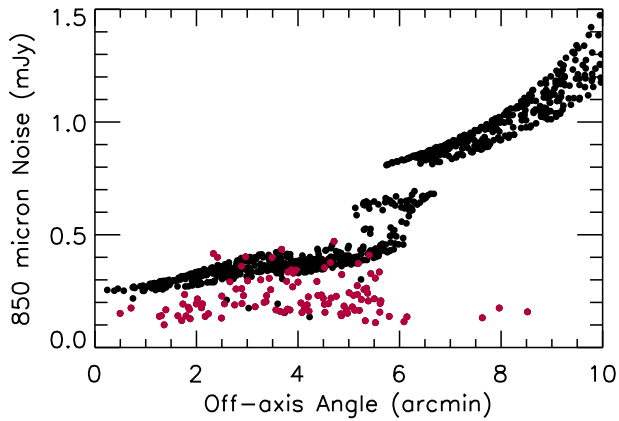}}
\caption{Locally determined 850~$\mu$m rms noise measurements 
for the sources in the full X-ray sample vs. off-axis angle 
(black circles---SCUBA-2 measurements, since these X-ray sources do not
lie in the ALMA observed areas; red circles---ALMA measurements).
The jump at $6'$ reflects the degrading sensitivity in the
off-axis SCUBA-2 images. It corresponds to the position interior to which the
SCUBA-2 maps are dominated by the more compact (DAISY) scan patterns.
(See C18 for an extensive discussion.)
\label{error_angle}
}
\end{figure}

For the X-ray sources that lie within the ALMA images but do not have
direct ALMA detections (i.e., there is no corresponding $>4.5\sigma$ 
submillimeter source), we measured the highest peak submillimeter flux 
within $1''$ of the X-ray source, along with the corresponding locally 
determined rms noise. 
This choice of radius is large relative to the measured dispersion between the
X-ray and ALMA positions (see Section~\ref{subsecxrayfluxes}).
In order to correct for the upward biasing introduced by the peaking 
up procedure, we applied a -0.41~mJy correction to each flux based on 
measurements made at random positions. 
We show the ALMA rms noise 
measurements as red circles in Figure~\ref{error_angle}.
We stress that we made all of the above measurements on primary beam
corrected images. We also did not make any measurements
outside the ALMA FWHM radius (see Section~\ref{subsecsubmm}).

We next subtracted from the SCUBA-2 image
all of the direct ALMA detections ($>4.5\sigma$) in
the ALMA images, smoothed to the SCUBA-2 
PSF (see Figure~8 of C18). This minimizes contamination 
from the wings of the bright submillimeter sources.
We then measured the SCUBA-2 fluxes and rms noise from the 
cleaned SCUBA-2 image for all of the remaining X-ray sources.  
We show the SCUBA-2 rms noise measurements as black circles in
Figure~\ref{error_angle}.

109 of the X-ray sources lie in the ALMA images with typical rms noise
$\sim0.1$--0.3~mJy. These sources are shown with
red circles in Figure 3.  The rms noise of the remaining X-ray sources measured 
from the SCUBA-2 cleaned image are higher, $\sim0.3$--0.5~mJy in the central 
region (see Figure~\ref{error_angle}). We note that the latter does not include confusion noise, 
which may be comparable to the systematic noise in this region (see C17).

If an X-ray source is detected at $>3\sigma$ at 850~$\mu$m, then we consider
the source to be submillimeter detected.
In the central region, 27 X-ray sources have submillimeter detections 
above 2.25~mJy and 37 above 1.65~mJy ($\sim7$\% of the central region 
X-ray sample). All of the sources with submillimeter fluxes above 2.25~mJy are ALMA fluxes,
while 4 of the 37 above 1.65~mJy are SCUBA-2 fluxes. Tests on randomized 
positions show that, on average, one of the detections above 1.65~mJy and 
none above 2.25~mJy will be false positives.

\subsection{Spectroscopic Redshifts and Spectral Classifications}
\label{subsecspec}
\begin{figure}[ht]
\centerline{\includegraphics[width=9cm,angle=0]{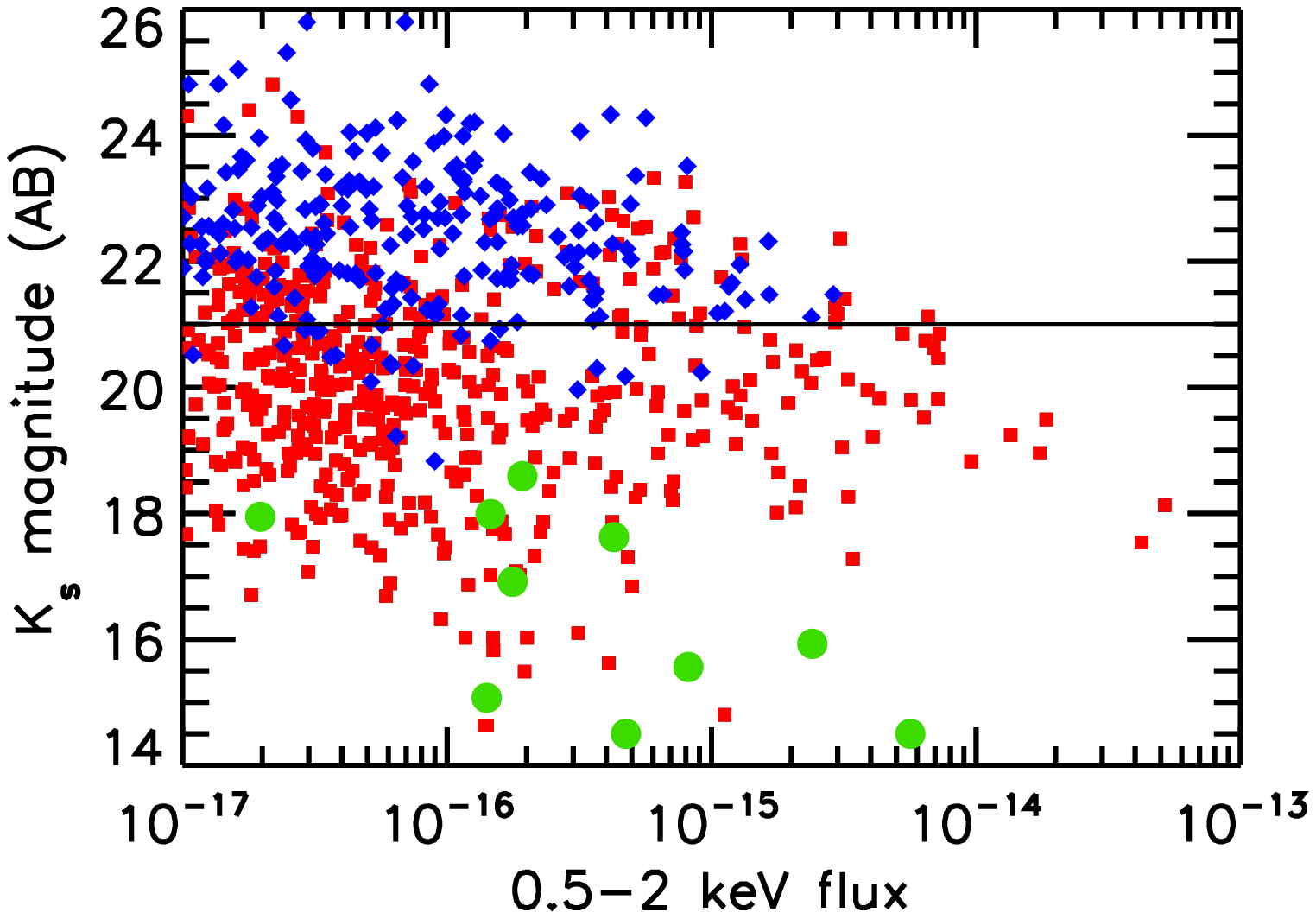}}
\centerline{\includegraphics[width=9cm,angle=0]{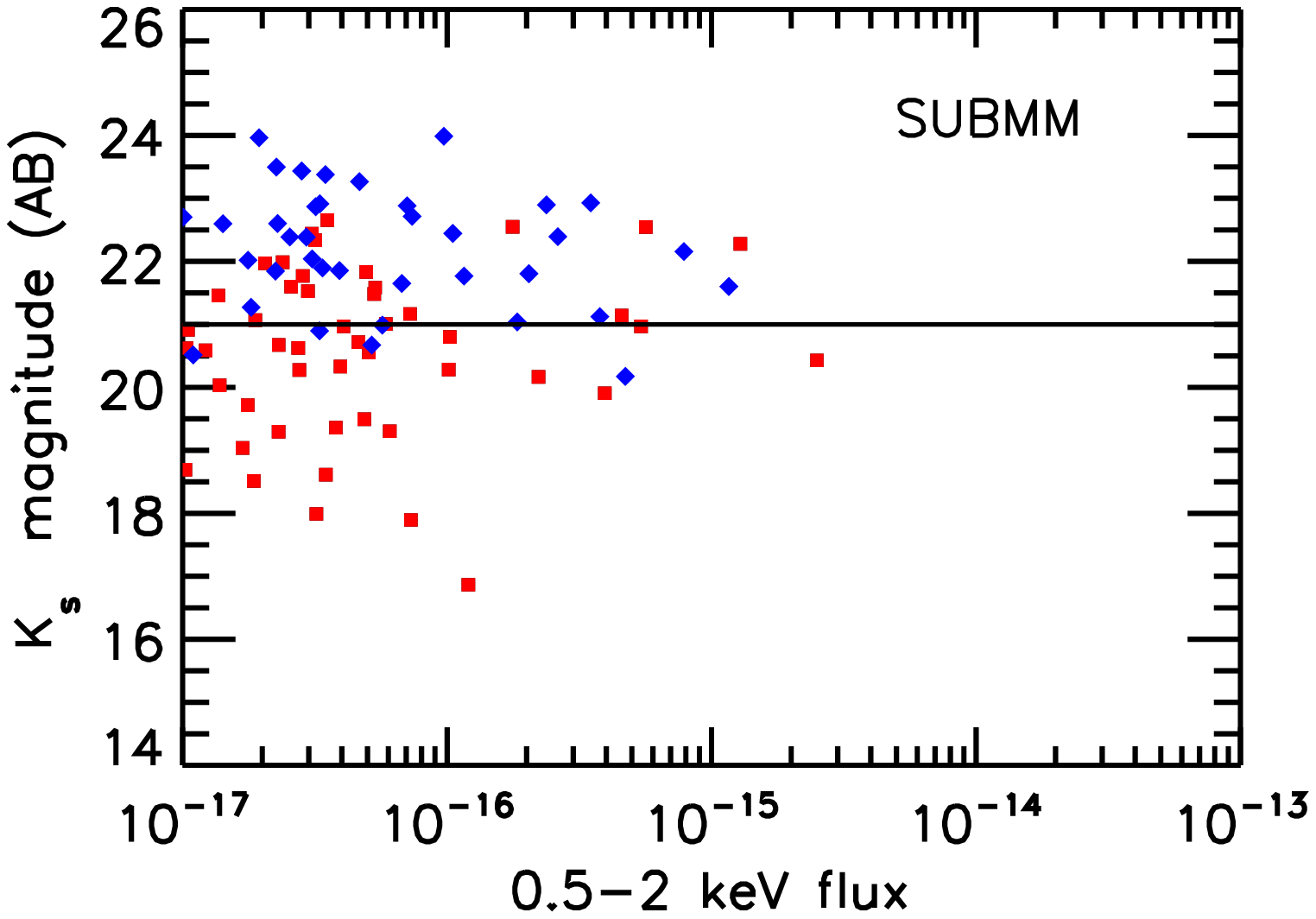}}
\caption{$K_s$ magnitude vs. observed 0.5--2~keV flux for (a) the full X-ray sample and
(b) those sources in the full X-ray sample that are detected at $>3\sigma$ at 
850~$\mu$m. Red squares denote X-ray sources with both a secure redshift and a
spectral classification (in (a), this includes 10 stars, denoted by
green circles), while the remaining sources are denoted by blue diamonds.
Brighter than $K_s=21$ (horizontal line), most of the
sources have both a secure redshift and a spectral 
classification: (a) 415 sources out of 442, or 94\%, and 
(b) 29 sources out of 34, or 85\%.
\label{fs_kmg_redshifts}
}
\end{figure}

L17 determined the most probable optical/NIR counterparts
to the X-ray sources based on both the $2\sigma$ positional uncertainties 
of the X-ray sources and the magnitudes of the potential counterparts. 
We adopt the Taiwan ECDF-S Near-Infrared Survey (TENIS) 
$K_s$ magnitudes (Hsieh et al.\ 2012) provided in the L17 catalog.
We show these magnitudes versus 
0.5--2~keV flux for the full X-ray sample in Figure~\ref{fs_kmg_redshifts}(a).

L17 next compiled redshifts from the literature for their most probable 
counterparts. From our full X-ray sample of 938 sources, L17 found 
529 (56\%) had redshifts considered secure by whichever 
group made the measurement. For a small
number of these, the X-ray source lies within the envelope
of a bright galaxy, but it is significantly separated from the galaxy's 
center. In these cases, we assumed that the X-ray source was associated
with the galaxy, and we assigned the redshift to the X-ray source. 
However, it is possible that the X-ray source is projected 
(i.e., an unrelated background source).

We compiled all of the publicly available spectra for these counterparts 
(see Popesso et al.\ 2009; Balestra et al.\ 2010; Cowie et al.\ 2012 and references therein;
see also the notes given in Table~1),
together with 377 spectra that we observed ourselves using the DEIMOS (Faber et al.\ 2003) 
(357 sources), LRIS (Oke et al.\ 1995) (13 sources) and MOSFIRE (McLean et al.\ 2012) (7 sources)
spectrographs on the Keck~10~m telescopes.
Details of how we reduced the spectra can be found in Cowie et al.\ (1996; 2016). 

We visually inspected the spectra for each counterpart and decided
whether the redshift identification was robust. We also
assigned a rough spectral class---broad-line AGN (BLAGN),
Seyfert type~2 (Sy2), absorber (Abs), and star-forming galaxy (SFG)---to each spectrum.
We classified as BLAGNs the sources that have some lines in their spectra
with FWHM\,$>2000$~km~s$^{-1}$.
Note that one object (L17~\#185) is a broad absorption line quasar, or BALQSO.
We classified as Sy2s the sources where high excitation narrow lines
are present (usually CIV~$\lambda1549$, CIII]~$\lambda1909$, or
[NeV]~$\lambda3426$; see, e.g., Szokoly et al.\ 2004).
We classified as SFGs the sources with UV absorption 
lines or EW([OII])$>10$~\AA\ and no broad or high-ionization lines, and as 
Abs the sources with no strong emission lines but strong absorption features.
The spectral classifications are subject to the lines that are visible in 
the available wavelength coverage and so may have some redshift 
dependence.

In this way, we assigned secure redshifts and spectral
classifications to 64\% of the sources in our full X-ray sample
(596 sources out of 938, including 10 stars).

We could not spectrally classify 17 of the sources in our full
X-ray sample that were cataloged in L17 as having secure redshifts. 
Based on the totality of the spectral data available,
we concluded that 10 of these do not have a secure redshift. 
For the remaining seven, either the spectra were not available to us, 
or they were insufficient for us to make a spectral classification, 
but we still adopted the redshift.
We also added a further two secure redshifts from the MOSDEF sample
of Kriek et al.\ (2015) for which we could not make spectral classifications.
Thus, in total, 605 sources in our full X-ray sample have 
secure redshifts (64.5\%), 596 of which also have spectral classifications.
We illustrate this in Figure~\ref{fs_kmg_redshifts}(a),
where we plot $K_s$ magnitude versus $0.5-2$~keV flux. 
We use red squares (or green circles for stars) to denote
X-ray sources with both a secure redshift and a spectral
classification, and we use blue diamonds to denote unclassified sources.

In contrast, in Figure~\ref{fs_kmg_redshifts}(b), we show how for the 
submillimeter detected X-ray sources, we only have
secure redshifts for 50\% (53 out of the 105 sources with 850~$\mu$m 
fluxes $>3\sigma$, of which 51 also have spectral classifications).
This presumably reflects that these sources are dustier, 
higher redshift, and have fainter $K_s$ magnitudes, which
makes them more difficult to identify with optical/NIR spectra.

We summarize the secure spectroscopic redshifts (hereafter, speczs)
and spectral classifications for the full X-ray sample in Table~1, where 
we also give the submillimeter fluxes and rms noise
(see Section~\ref{subsecsubmmfluxes}). 
In the full X-ray sample, 
there are 22 BLAGNs, none of which is detected
at $>3\sigma$ at 850~$\mu$m (the one BALQSO is also not detected);
47 Sy2s, 8 of which are detected; 388 SFGs, 38 of which are detected; 
and 128 Abs, 5 of which are detected.
The largest number of submillimeter detected sources
occurs in the category without spectral classifications
(no IDs; 54 out of 342), as might be expected given the higher
obscuration and higher redshifts of submillimeter sources
in general.

\begin{figure}[ht]
\centerline{\includegraphics[width=9cm,angle=0]{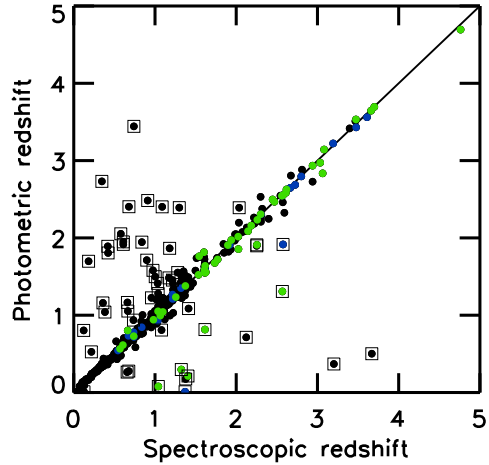}}
\caption{Adopted photzs from H14
and S16 vs. secure speczs for sources in the full X-ray sample 
with both a specz and a photz (black circles). 
The BLAGNs are shown in blue, and the Sy2s in green. 
Outliers are enclosed in open squares.
\label{compare_photz}
}
\end{figure}

\subsection{Photometric Redshifts}
\label{subsecphotz}
There has been a large number of papers estimating photometric
redshifts (hereafter, photzs) in the CDF-S 
(e.g., Santini et al.\ 2009, 2015; Rafferty et al.\ 2011; 
Dahlen et al.\ 2013; Hsu et al.\ 2014; Skelton et al.\ 2014; Straatman et al.\ 2016).
Rather than recompute photzs ourselves, we use 
the literature estimates here. However, we caution that in C18, 
we found considerable scatter in the estimates from the different
literature catalogs for the optical/NIR faint submillimeter sources.

We compared the photzs from the literature catalogs that use the
deep {\em Spitzer\/} IRAC data of either Ashby et al.\ (2013) or
Labb{\'e} et al.\ (2015) and are purely photometrically based
(i.e., Hsu et al.\ 2014, hereafter, H14; Santini et al.\ 2015;
Straatman et al.\ 2016, hereafter, S16) against the 
speczs in Table~1. We found that they each have a significant number of 
outliers, which we define as (photz-specz)/specz$> 0.1$.
(Note that our outlier definition is tighter than that of L17, who used $>0.15$.)
All three catalogs have 9--10\% of outliers, but H14
has a much lower percentage of outliers than the other two catalogs 
when considering only the X-ray sources with AGN spectral signatures. 
Thus, we adopt the photzs first from H14, and if they do not have one, then 
from S16 (after restricting to sources with their quality flag $Q<3$). All but 44 
of the X-ray sources have either speczs or photzs, and of those 44, 
all but one are faint ($K_s>25$). The brighter source is closely blended 
with a brighter galaxy.

In Column~12 of Tables~1 and 2, we give the adopted photzs for the 
full X-ray sample and the ALMA sample, respectively.
Additionally, in Table~2 we provide a FIR-based redshift estimate from C18 in brackets 
if there is no specz or high-quality photz available.
In Figure~\ref{compare_photz}, we compare the photzs and speczs
for sources in the full X-ray sample with both a specz and a photz (black circles).
We mark the outliers by enclosing them in open squares.
The photzs for the BLAGNs and the Sy2s (blue and green, respectively) 
are generally well estimated. 

There are only a handful of X-ray sources with high redshifts
(nine sources above $z=4$ and two above $z=5$), almost
all of which are based on photzs (see Cowie et al.\ 2019 for
a detailed discussion of these very high-redshift AGN candidates). 
Most of the X-ray sources lie at low redshifts, with 570 (58\%) 
below $z=1.6$. Most of these low-redshift sources 
have spectroscopic identifications (501 sources).

\subsection{X-ray Luminosities}
\label{subsecxraylum}
Given the high redshifts of the submillimeter detected sources
(see Figure~25 of C18), 
we calculate the rest-frame X-ray luminosities from
\begin{equation}
L^R_{\rm 2-8~keV} = 4\pi d_L^2 f_{\rm 0.5-2~keV} ((1+z)/4)^{\Gamma-2}~{\rm erg~s^{-1}} \,,
\end{equation}
and 
\begin{equation}
L^R_{\rm 8-28~keV} = 4\pi d_L^2 f_{\rm 2-7~keV} ((1+z)/4)^{\Gamma-2}~{\rm erg~s^{-1}} \,,
\end{equation}
where $d_L$ is the luminosity distance, and we take the photon index $\Gamma=1.8$.
These equations are exact for sources at $z=3$, but the K-correction term may be less
appropriate for lower redshift sources.
We have not corrected the luminosities for X-ray absorption, which could potentially have
a non-negligible effect, particularly for the lower energy band. We will consider
absorption corrected luminosities in the Discussion.

\section{850$~\mu\lowercase{m}$ Properties of the X-ray Sample}
\label{sec850props}
The 850~$\mu$m flux is a rough measure of the FIR luminosity
and hence of the SFR, independent of redshift (Blain \& Longair 1993).
In rough terms, a 2~mJy source corresponds to a 
${\rm SFR} \approx 300\ {\rm M}_\odot~{\rm yr}^{-1}$ for a Kroupa (2001)
initial mass function (IMF); see Equation~5 of C17, which gives 
\begin{equation}
{\rm SFR}\, ({\rm M}_\odot\,{\rm yr}^{-1})=(143 \pm 20) \times S_{850\,\mu {\rm m}}\, ({\rm mJy}) \,.
\end{equation}
This SFR is not usually affected by AGN contributions, since
the torus is generally too hot to contribute to the $850~\mu$m
flux, even at high redshifts (e.g., Hatziminaoglou et al.\ 2010).

\subsection{850$~\mu\lowercase{m}$ Signal}
\label{subsecsignal}

\begin{figure}[ht]
\centerline{\includegraphics[width=9cm,angle=0]{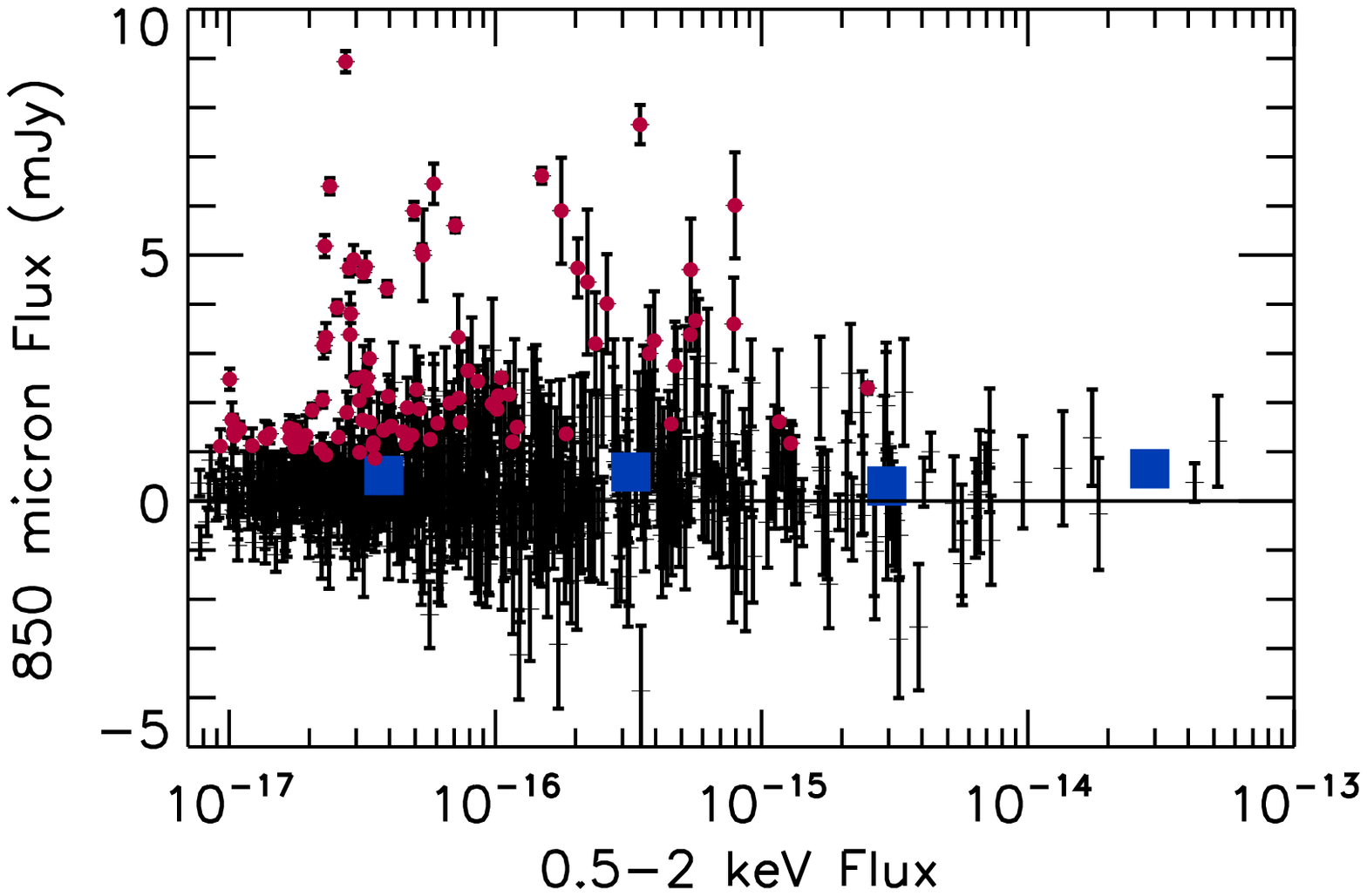}}
\centerline{\includegraphics[width=9cm,angle=0]{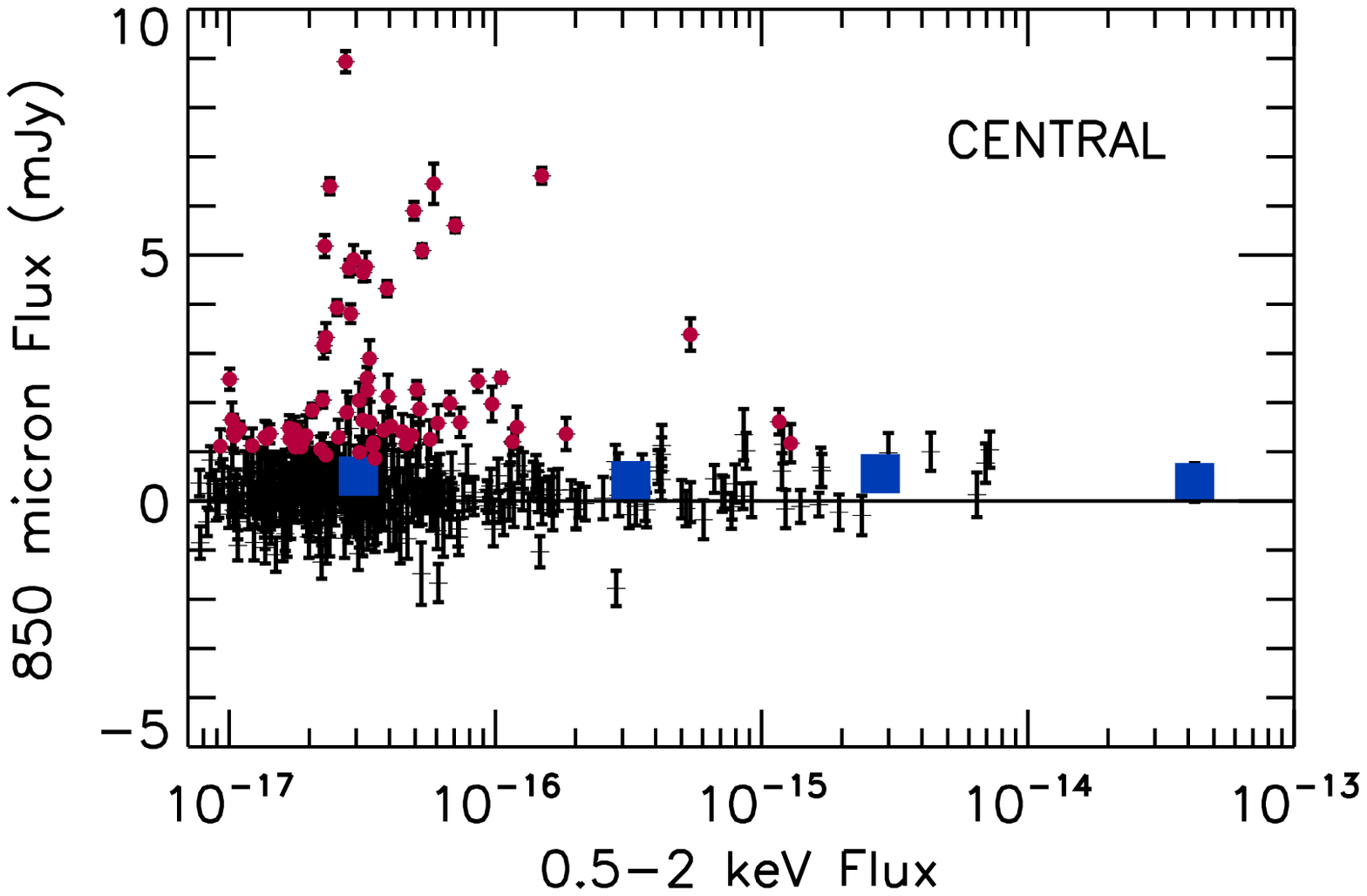}}
\caption{Measured 850~$\mu$m flux vs. 0.5--2~keV flux
(black with $1\sigma$ uncertainties) for the (a) full 
and (b) central region X-ray samples.
The red circles show the sources with $>3\sigma$ 850~$\mu$m detections.
The blue squares show the mean 850~$\mu$m flux per X-ray source
(here the uncertainties are generally smaller than the symbol size).
\label{flux_smm}
}
\end{figure}

In Figure~\ref{flux_smm}(a), we show measured 850~$\mu$m flux versus
0.5--2~keV flux for the full X-ray sample. The X-ray sources contain a 
significant 850~$\mu$m flux with a mean 
of 0.52~mJy per X-ray source, or just over
a ${\rm SFR}$ of $70~{\rm M}_\odot~{\rm yr}^{-1}$.
The 95\% confidence range is 0.44 to 0.61~mJy.
As can be seen from the blue squares, this contribution
is similar at all X-ray fluxes. We also find a similar value for the
central region X-ray sample (Figure~\ref{flux_smm}(b)), where both 
the X-ray and submillimeter data are much more sensitive. 
Here the mean 850~$\mu$m flux is 0.49~mJy 
per X-ray source, with a 95\% confidence range of 0.39--0.60~mJy.
However, a Kolmogorov-Smirnov test gives a less than 1\% probability 
that either of the distributions is normal, and the medians are substantially 
lower: 0.26~mJy for the full sample, and 0.22~mJy for the central sample.

\begin{figure}[ht]
\centerline{\includegraphics[width=9cm,angle=0]{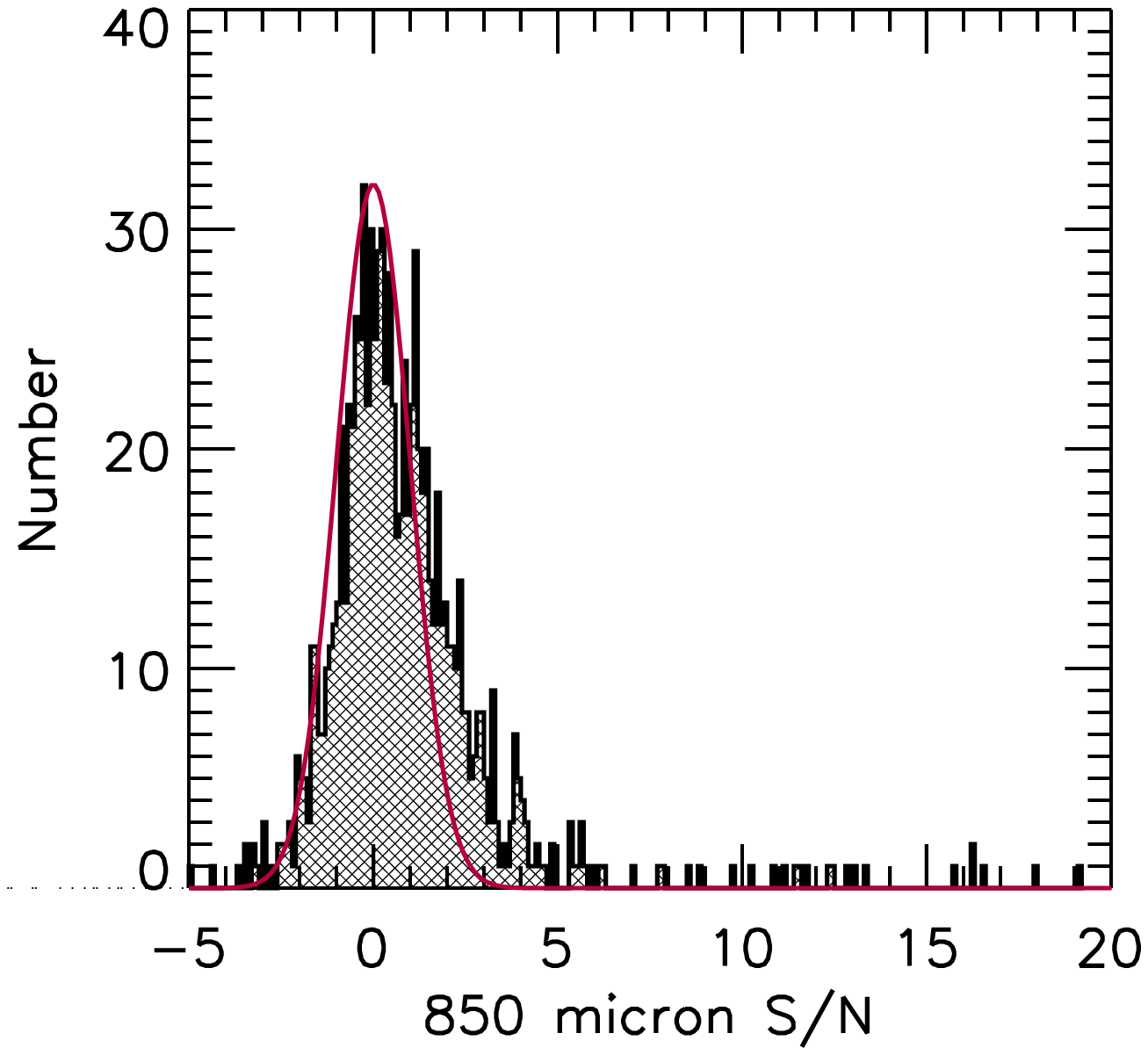}}
\centerline{\includegraphics[width=9cm,angle=0]{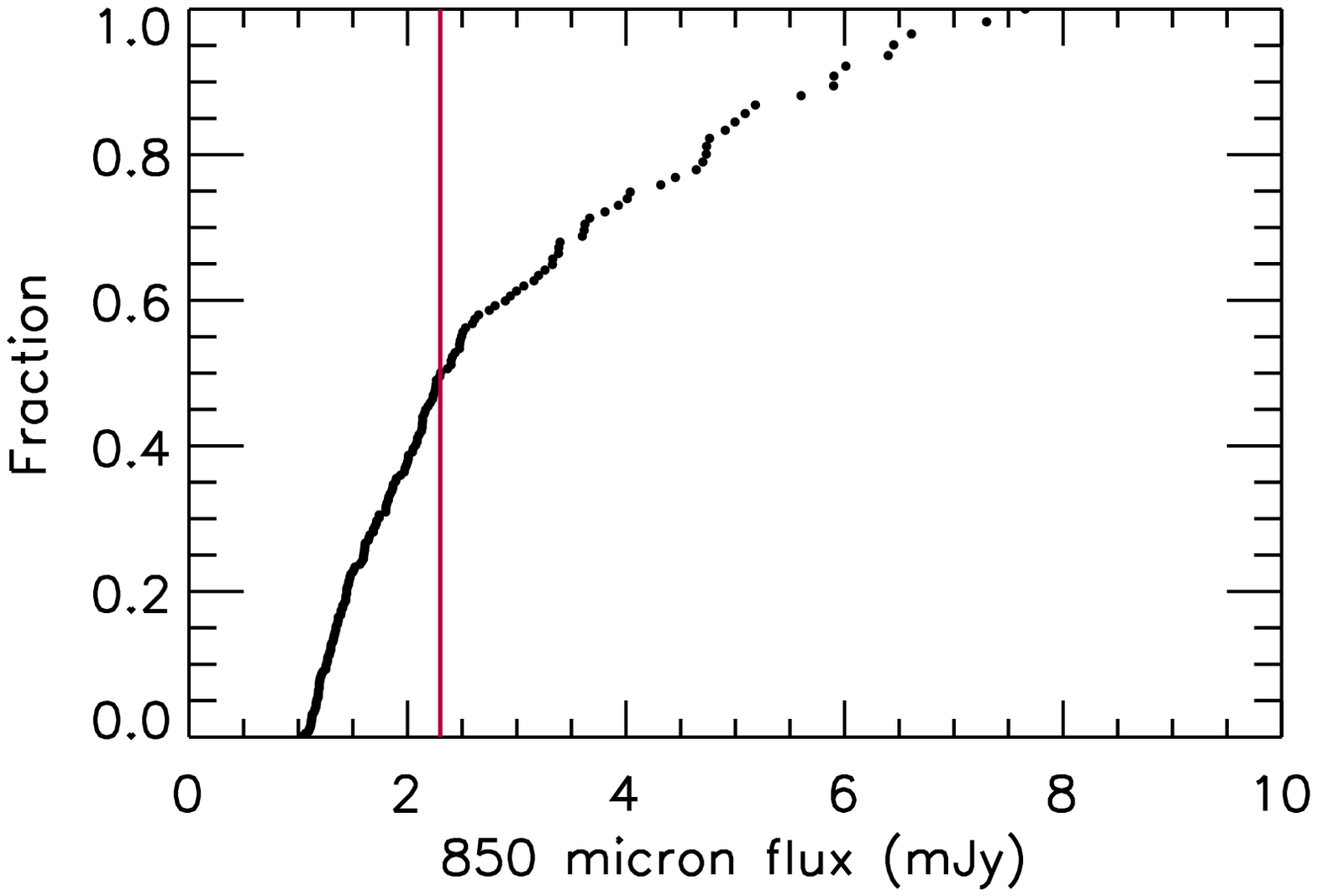}}
\caption{(a) S/N distribution for the 850~$\mu$m measurements
of the full X-ray sample (gray shading).
The red curve shows the expected distribution in the absence
of any submillimeter signal in the X-ray sample.
(b) Cumulative contribution with increasing 850~$\mu$m flux 
to the total measured 850~$\mu$m light from the full X-ray sample. 
50\% of the light comes from the 57 (out of 938) sources with fluxes 
above 2.3~mJy (red vertical line).
\label{smm_hist}
}
\end{figure}

Nearly all of the 850~$\mu$m signal comes from the sources
that are significantly detected.
Barger et al.\ (2015) found a similar result for the GOODS-N.
In Figure~\ref{smm_hist}(a), we show the S/N distribution 
(gray shading), which illustrates the skewness and kurtosis of the
measurements. 
The bulk of the X-ray sources are consistent with having no 850~$\mu$m
flux. The red curve shows the expected distribution in this case.
The sources producing most of the 850~$\mu$m flux lie
in the very extended tail. This illustrates the dangers of
using simple stacking analyses on this type of data, where
a very small number of sources dominate the mean and are
poorly representative of the great majority of the sample.
To emphasize this point,
in Figure~\ref{smm_hist}(b), we show the cumulative contribution
to the total measured 850~$\mu$m light from the full X-ray sample.
50\% of the light comes from the 57 (out of 938) sources with fluxes 
above 2.3~mJy.

\subsection{Contributions to the 850$~\mu\lowercase{m}$ Extragalactic Background Light}
\label{subsecEBL}

\begin{figure}[ht]
\centerline{\includegraphics[width=9cm,angle=0]{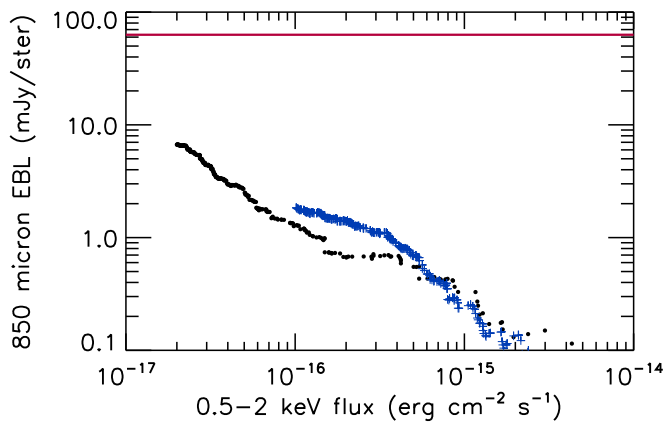}}
\caption{850~$\mu$m EBL produced by the full (blue crosses;
down to the 0.5--2~keV completeness level of 
$10^{-16}$~erg~cm$^{-2}$~s$^{-1}$)
and central region (black circles; down to the deeper 0.5--2~keV
completeness level of $2\times10^{-17}$~erg~cm$^{-2}$~s$^{-1}$) 
X-ray samples vs. 0.5--2~keV flux. 
The red horizontal line shows the 850~$\mu$m EBL from 
Fixsen et al.\ (1998). Because
there are more bright X-ray sources in the full X-ray sample,
it is a better estimator of the contributions to the EBL
from the brighter X-ray fluxes, while the deeper central X-ray sample is a
better estimator of the contributions from the fainter X-ray fluxes.
\label{ebl_smm}
}
\end{figure}

In Figure~\ref{ebl_smm}, we show the contributions 
above a given 0.5--2~keV flux to the 850~$\mu$m extragalactic 
background light (EBL) from the full X-ray sample down to its completeness 
level of $10^{-16}$~erg~cm$^{-2}$~s$^{-1}$ (blue crosses) and 
from the central region X-ray sample down to its deeper completeness 
level of $2\times10^{-17}$~erg~cm$^{-2}$~s$^{-1}$ (black circles).
The contributions rise rapidly with decreasing X-ray flux. 
At the limits of the {\em Chandra\/} 7~Ms data, $\sim10$\% of the 
850~$\mu$m EBL measured by Fixsen et al.\ (1998) is identified by 
the X-ray sources (which encompasses more than just AGNs, see below). 
Deep blank-field SCUBA-2 surveys recover $\sim30$\%
of the Fixsen et al.\ 850~$\mu$m EBL measurement 
(e.g., Zavala et al.\ 2017), which means most of that submm light is 
not being identified by the X-ray sources, even though the sensitivity
of the central region of the CDF-S data is such 
that we could detect Compton-thin X-ray 
AGNs out to beyond $z\sim6$ (see Figure~1 of Cowie et al.\ 2019).
As we discuss below, some of the X-ray light in
the submillimeter detected X-ray sources is due to star formation rather
than AGN activity, so this constraint is even more severe.

\subsection{Star-Forming Galaxies versus AGNs}
\label{subsecAGNSF}
We preface this section by emphasizing that we have not corrected
our X-ray luminosities for absorption (see Section~\ref{subsecxraylum}),
which could cause some uncertainties in the luminosity bins
that we adopt.  However, by focusing on a
2--7~keV selected sample, we will be less sensitive to 
obscuration effects for Compton-thin AGNs than we would be
with a 0.5--2~keV selected sample, though we need to keep in mind
that even a 2--7~keV selected sample will not be able to probe 
very Compton-thick AGNs.

\begin{figure}
\centerline{\includegraphics[width=9cm,angle=0]{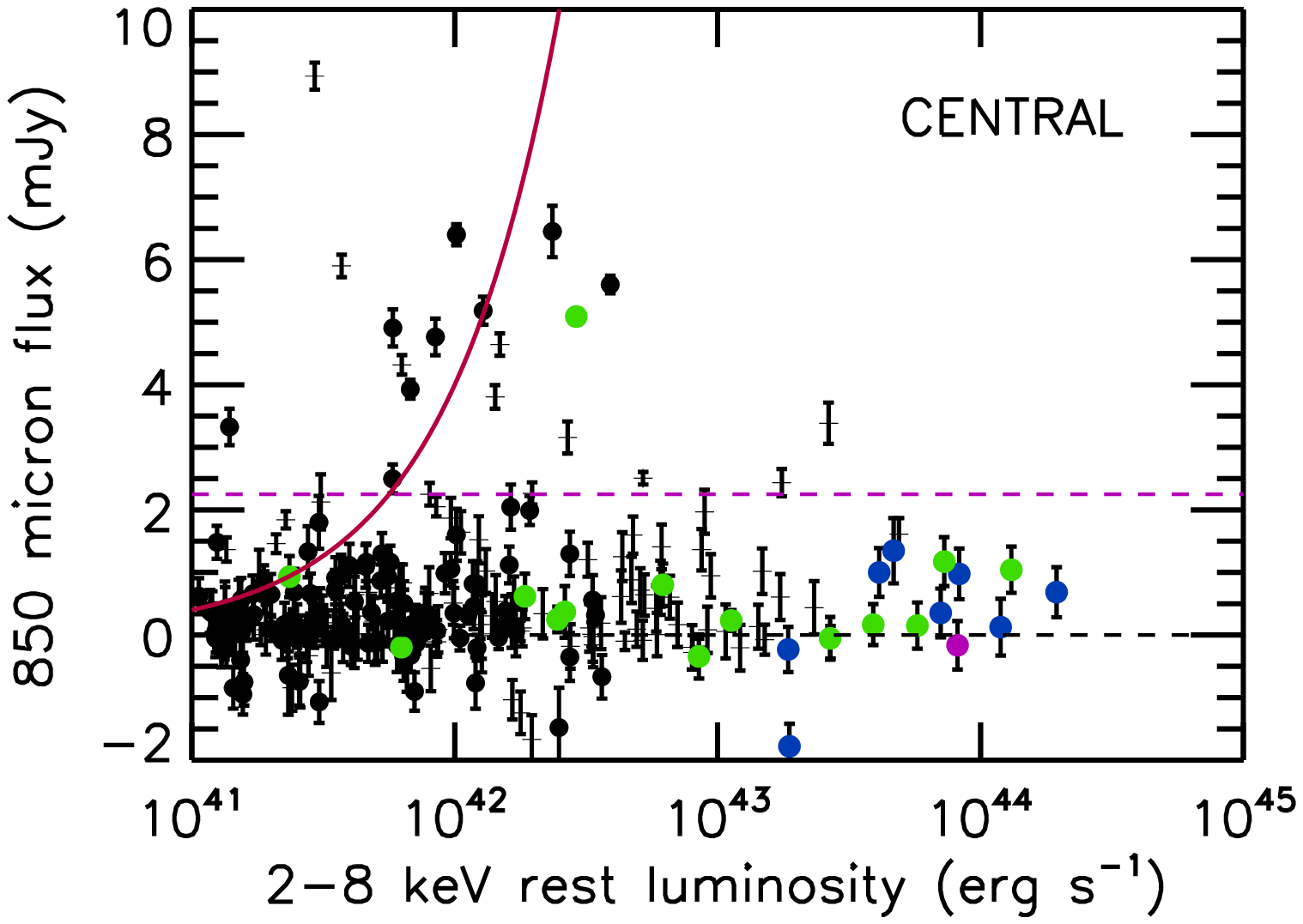}}
\centerline{\includegraphics[width=9cm,angle=0]{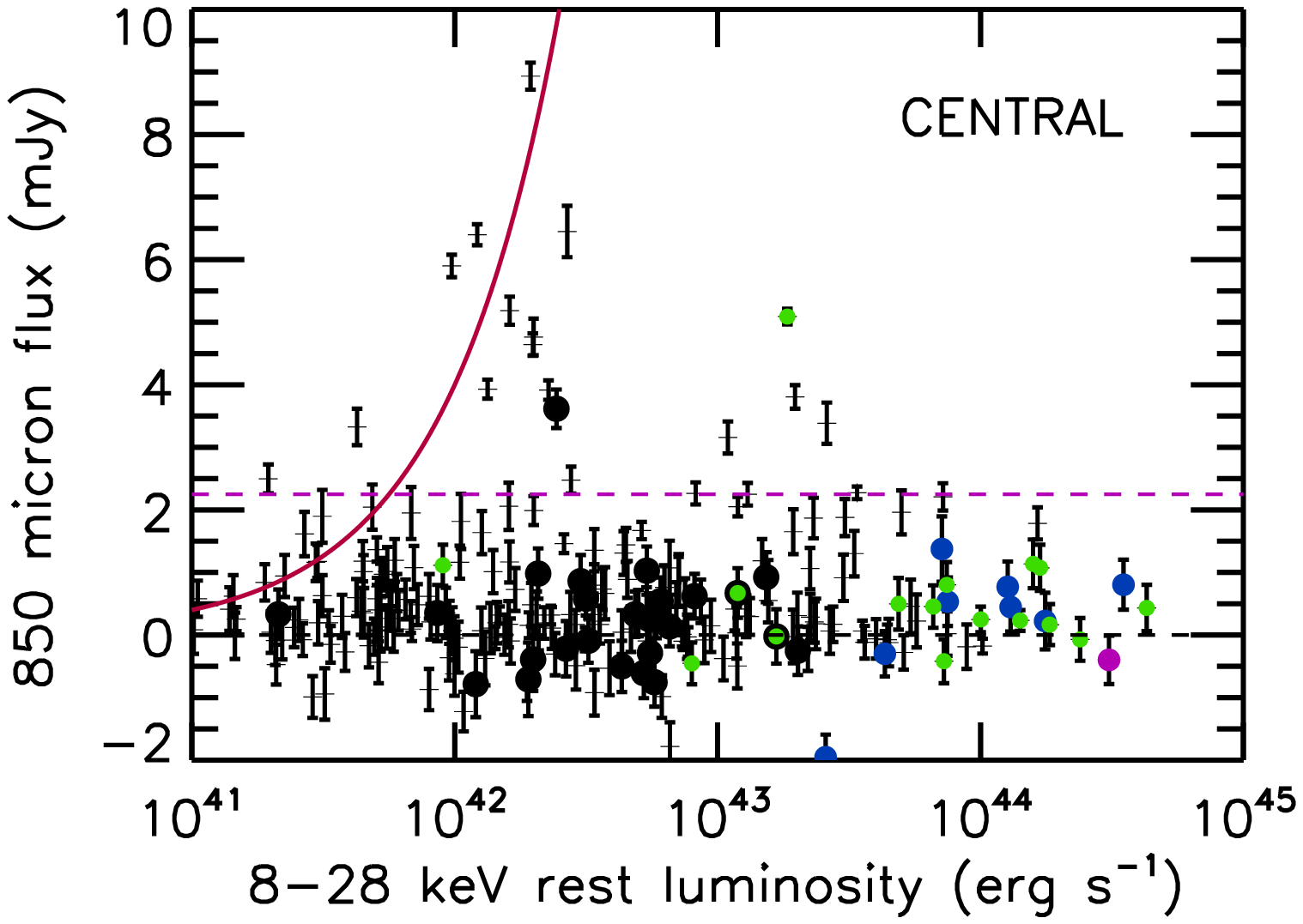}}
\caption{850~$\mu$m flux vs. (a) $L^R_{\rm 2-8~keV}$
and (b) $L^R_{\rm 8-28~keV}$
for the central region X-ray sample with $z>1$ (specz and photz).
Blue circles show BLAGNs, green 
Sy2s, and purple the one BALQSO. 
All other sources are shown as black plus signs.
The uncertainties are $\pm1\sigma$.
The red curve shows the relation for sources where the 
X-rays are due to star formation.
Sources not detected in the other X-ray band are shown
with black circles. In (a) there are no Sy2s or BLAGNs that
satisfy this, while in (b) only two Sy2s do, and they are shown 
as black circles with green interiors.
The purple dashed horizontal line in each panel shows the 
2.25~mJy 850~$\mu$m limit, which roughly corresponds to the 
SCUBA-2 $>4\sigma$ limit through the whole central region.
\label{lx_smm}
}
\end{figure}

In Figure~\ref{lx_smm}, we show 
850~$\mu$m flux versus (a) $L^R_{\rm 2-8~keV}$ and 
(b) $L^R_{\rm 8-28~keV}$ for the central region X-ray sample.
We only show sources with $z>1$ 
(specz and photz), since for these redshifts, the 850~$\mu$m flux
is a good measure of the FIR luminosity. Sources that are only
detected in one X-ray band and hence are only present in one
panel are shown with black circles (or, in the case of the two Sy2s
in Figure~\ref{lx_smm}(b), black circles with green interiors).
We obtained the star formation locus (red curve in each panel), where 
both the X-ray luminosity and the 850~$\mu$m flux are consistent with 
being produced by star formation, using
Mineo et al.\ (2012)'s $L_X$-SFR relation (from their Section~8.1, but for 
a Kroupa 2001 IMF and our X-ray luminosities) and 
C17's SFR-$f_{\rm 850~\mu m}$ relation (their Equation~5, which we
reproduce as Equation~3 in this paper).

The brightest 850~$\mu$m detected X-ray sources, most
of which are only detected in the more sensitive 0.5--2~keV band 
(i.e., black circles in Figure~\ref{lx_smm}(a)), are drawn primarily 
from low X-ray luminosity galaxies.
Considering both panels and taking into account expected
scatter about the star formation 
locus, we conclude that all but one of the sources with high 850~$\mu$m 
fluxes ($>4$~mJy) are consistent with 
having both their X-ray luminosity and their 850~$\mu$m flux produced
by star formation. The exception is L17~\#666: this Sy2 has an 850~$\mu$m flux
of $5.1\pm0.13$~mJy and an 
$L^{R}_{\rm 8-28~keV} = 1.8\times10^{43}$~erg~s$^{-1}$, and
because it is faint in the 0.5--2~keV band but bright in the 2--8~keV band,
it is likely a highly obscured AGN.

There are only eight sources that are 
strong submillimeter sources (850~$\mu$m fluxes $>2.25$~mJy, which
is roughly the SCUBA-2 $>4\sigma$ limit through the whole central region)
and have X-ray luminosities that clearly classify them as AGNs
(here, $L^{R}_{8-28~{\rm keV}} > 10^{42.5}$~erg~s$^{-1}$).
However, these sources are all at intermediate (Seyfert) X-ray luminosities 
($L^{R}_{8-28~{\rm keV}} = 10^{42.5}$--$10^{44}$~erg~s$^{-1}$) with
a range from $L^{R}_{8-28~{\rm keV}} = 0.8$--$7.2\times10^{43}$~erg~s$^{-1}$;
none of the 19 sources at high (quasar) X-ray luminosities 
($L^{R}_{8-28~{\rm keV}} > 10^{44}$~erg~s$^{-1}$)
has an 850~$\mu$m flux above 1.6~mJy. Indeed, the mean 850~$\mu$m 
flux for these 19 host galaxies is only 0.46~mJy, which 
corresponds to a SFR of $\sim 65~{\rm M}_\odot~{\rm yr}^{-1}$.
We conclude that extreme SFRs $\gtrsim300~{\rm M}_\odot~{\rm yr}^{-1}$ 
are only seen at intermediate X-ray luminosities.

\begin{figure}
\centerline{\includegraphics[width=9cm,angle=0]{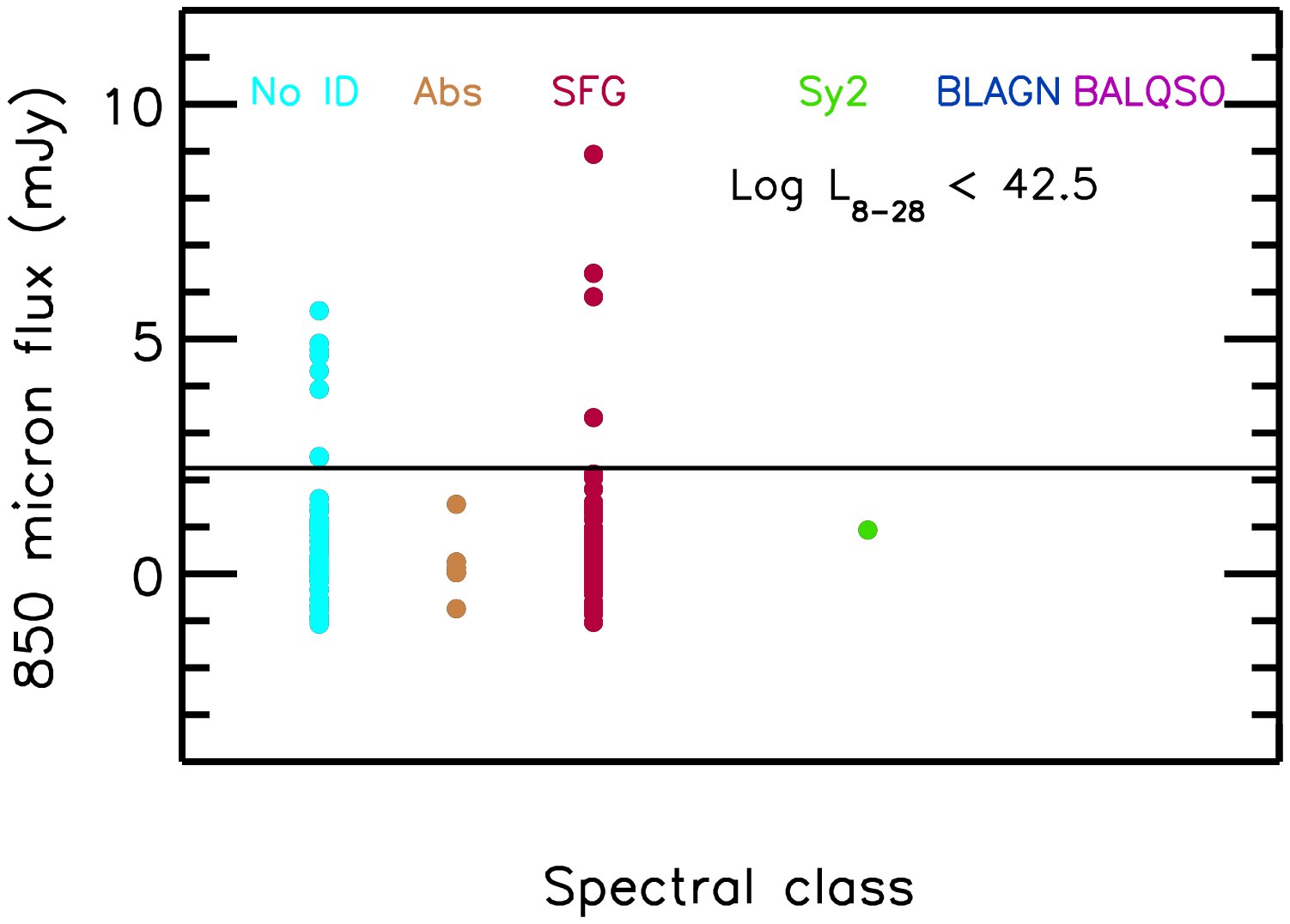}}
\centerline{\includegraphics[width=9cm,angle=0]{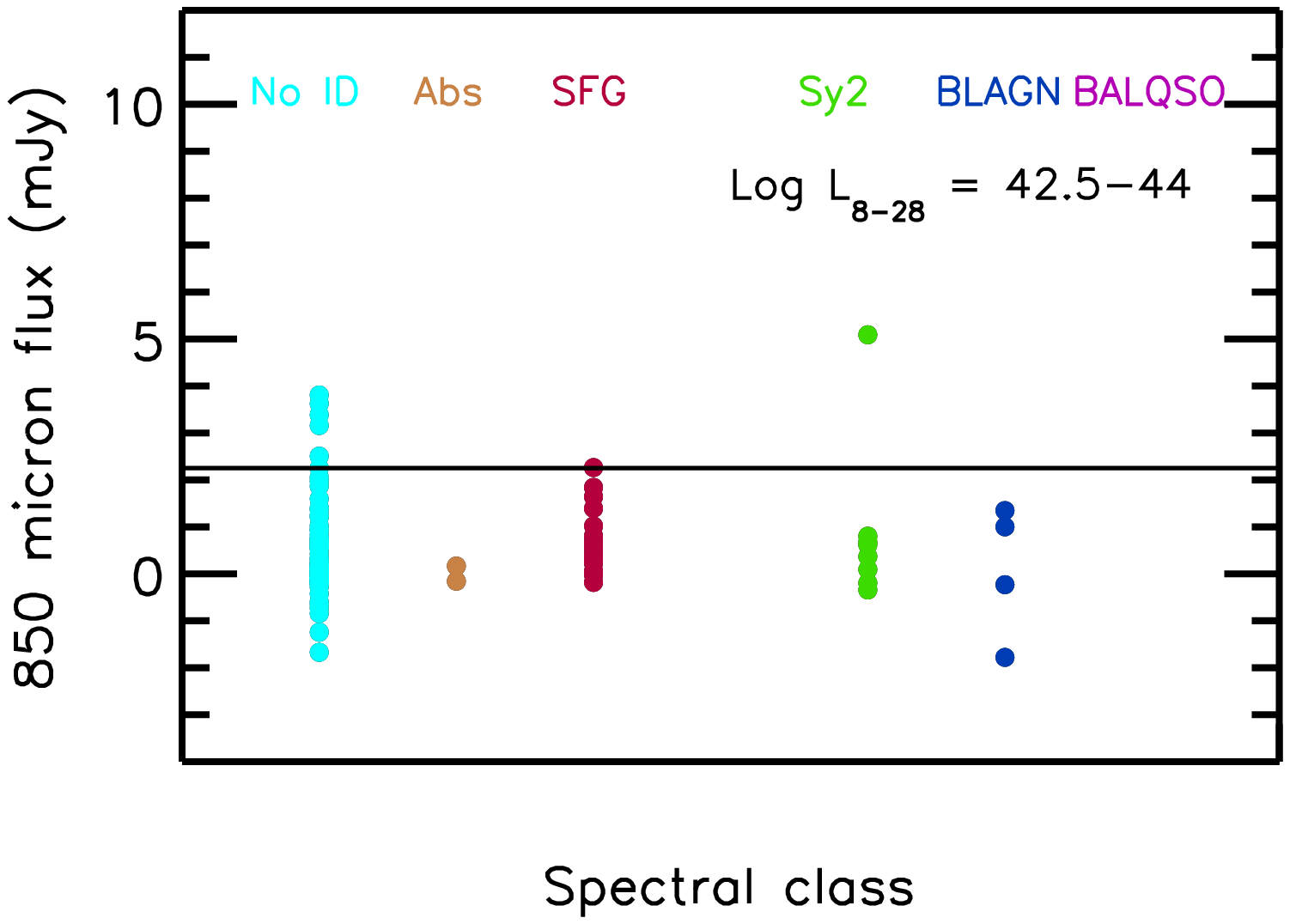}}
\centerline{\includegraphics[width=9cm,angle=0]{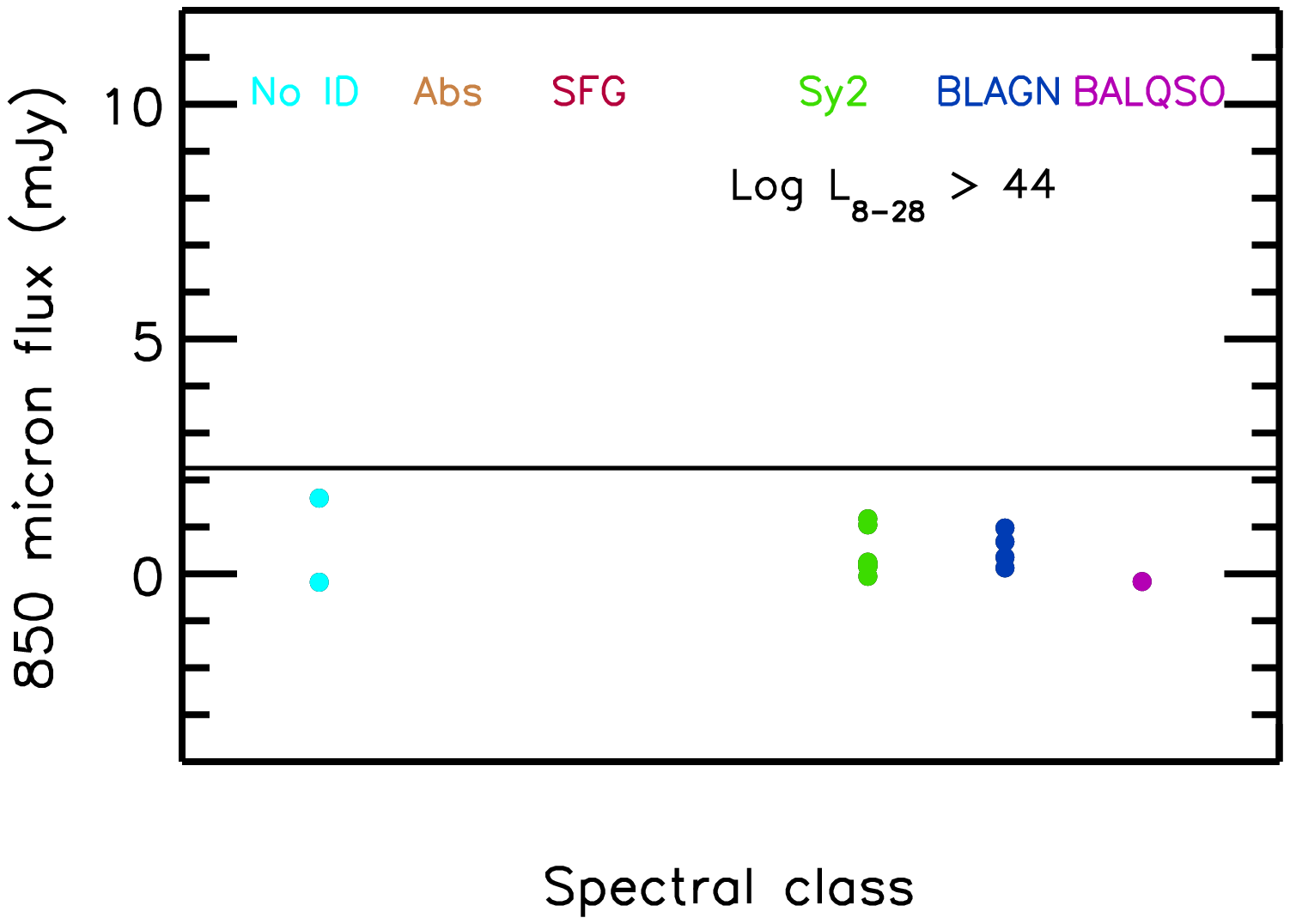}}
\caption{850~$\mu$m flux vs. optical spectral class
for the central region X-ray sample with $z>1$ (specz and photz)
and (a) $\log L^{R}_{8-28~{\rm keV}}<42.5$, 
(b) $\log L^{R}_{8-28~{\rm keV}}=42.5$--44, and
(c) $\log L^{R}_{8-28~{\rm keV}}>44$.
The optical spectral classifications are color-coded (purple---BALQSO;
blue---BLAGNs; green---Sy2s; red---SFGs; gold---Abs; cyan---no IDs).
The horizontal line shows the 2.25~mJy 850~$\mu$m limit, which
roughly corresponds to the SCUBA-2 $>4\sigma$ limit 
through the whole central region.
\label{flux_class}
}
\end{figure}

In Figure~\ref{lx_smm}(b), we can see that the decline in
850~$\mu$m fluxes of individual sources above
$\sim5\times 10^{43}$~erg~s$^{-1}$
corresponds roughly to the X-ray luminosity where
the BLAGNs and many of the Sy2s begin to appear. We illustrate this more
clearly in Figure~\ref{flux_class}, where we plot 850~$\mu$m flux versus spectral
class for three bins in X-ray luminosity. The most X-ray luminous (quasar) bin 
is dominated by BLAGNs and Sy2s, and none of the sources
has an 850~$\mu$m flux $>2.25$~mJy. In the intermediate X-ray luminosity bin, 
there is a much wider range of spectral classes, and a fraction of the sources 
(though none of the BLAGNs) have 850~$\mu$m fluxes $>2.25$~mJy.
In the lowest X-ray luminosity bin, SFGs and no IDs dominate,
and a significant fraction of these have 850~$\mu$m fluxes $>2.25$~mJy. 
For these sources, the X-ray luminosity is likely produced by star formation.

These results suggest that the strongest star formation occurs in less luminous 
X-ray AGNs, while extreme star formation is not seen in the most luminous 
X-ray AGNs.  We quantify and interpret this result in the Discussion.

\section{X-ray properties of the Submillimeter Detected X-ray Sources}
\label{secgamma}
We now turn to the X-ray properties of the submillimeter detected 
X-ray sources, and, in particular, to whether the X-ray spectra 
are consistent with the classifications
we inferred from Figure~\ref{lx_smm}. As in
Section~\ref{subsecAGNSF}, we focus on the central 
region, where the X-ray and submillimeter observations are the deepest.

\begin{figure}[ht]
\centerline{\includegraphics[width=9cm,angle=0]{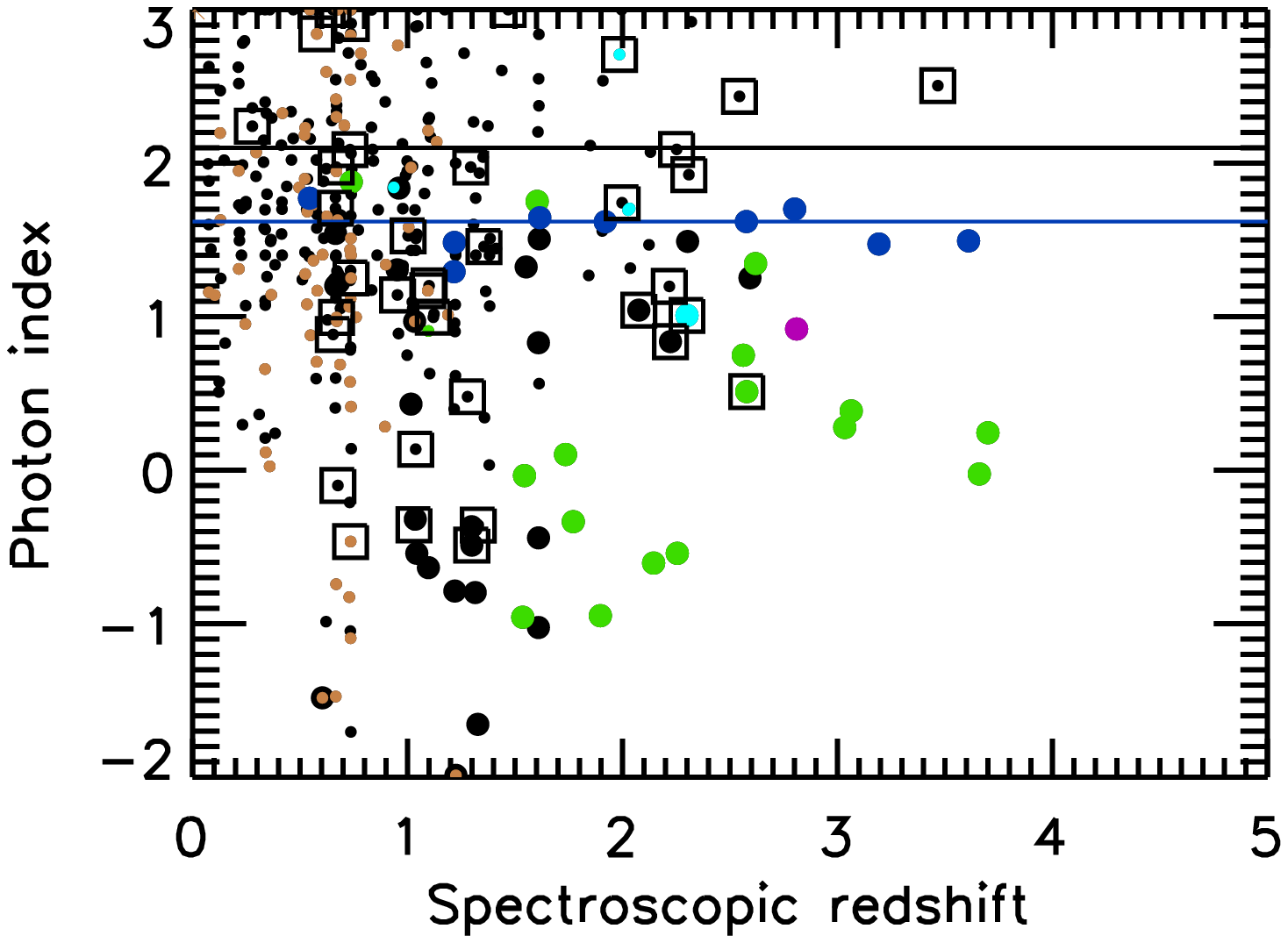}}
\centerline{\includegraphics[width=9cm,angle=0]{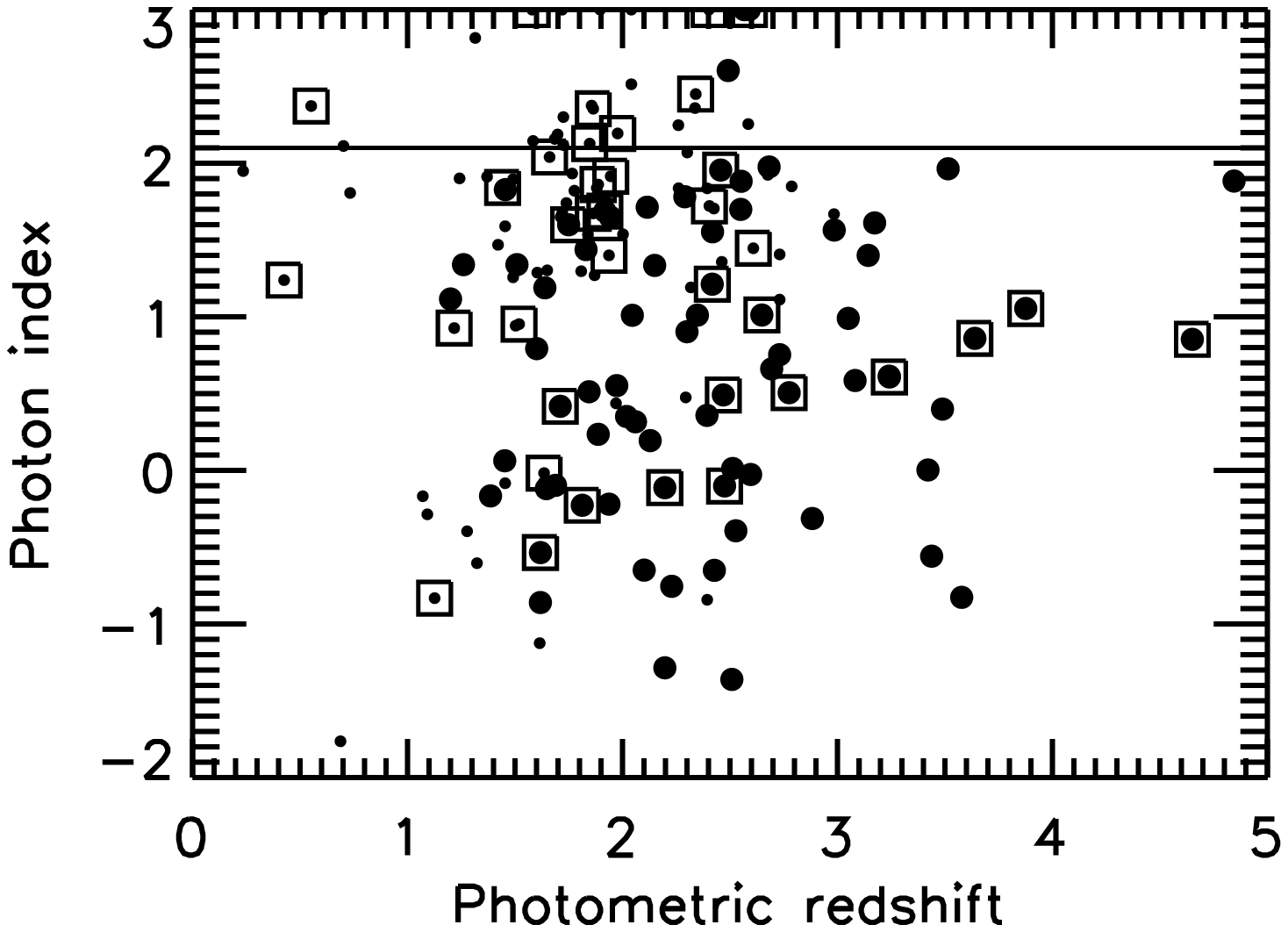}}
\caption{(a) Effective observed photon index from L17 vs. redshift 
for the central region X-ray sample with speczs. 
The blue line shows $\bar{\Gamma}=1.62$ for the BLAGNs. 
The optical spectral classifications are color-coded (purple---BALQSO;
blue---BLAGNs; green---Sy2s; black---SFGs; gold---Abs; cyan---no IDs).
(b) The same, but for the central region X-ray sample with only photzs.
There are no spectral classifications for these sources, so they are all
shown as black. In both panels,
sources with $L^{R}_{8-28~{\rm keV}} > 10^{42.5}$~erg~s$^{-1}$
are shown with larger symbols, sources with $>3\sigma$ detections 
at 850~$\mu$m are enclosed in open squares, and $\Gamma=2.1\pm0.1$ for
luminous, high-mass X-ray binaries from Sazonov \& Khabibullin (2017)
is indicated with the black line. 
\label{gamma}
}
\end{figure}

We start by considering the properties of the central region 
X-ray sample with speczs. 
In Figure~\ref{gamma}(a), we plot effective observed photon index,
$\Gamma$, from L17 versus specz, color-coding the sources by 
optical spectral class. We show the sources with
$L^{R}_{8-28~{\rm keV}} > 10^{42.5}$~erg~s$^{-1}$
with larger symbols. We enclose in an open square any source with 
a $>3\sigma$ submillimeter detection. 

Only one of the sources classified as an AGN from the optical/NIR 
spectra---the Sy2 galaxy L17~\#666 
discussed in Section~\ref{subsecAGNSF}---has a submillimeter detection 
(green circle enclosed in an open square in Figure~\ref{gamma}(a)).
Note that the BLAGNs (blue) all have nearly identical photon indices:  
in the central region, the mean photon index is $\bar{\Gamma}=1.62$
(blue line in Figure~\ref{gamma}(a)), and in the full sample,
$\bar{\Gamma}=1.65$. Their uniformity is consistent with there being very 
little absorption in their X-ray spectra. In contrast, the Sy2s (green) have
a wide range in $\Gamma$ corresponding to a large spread in the 
absorption column densities.

\begin{figure}[ht]
\centerline{\includegraphics[width=9cm,angle=0]{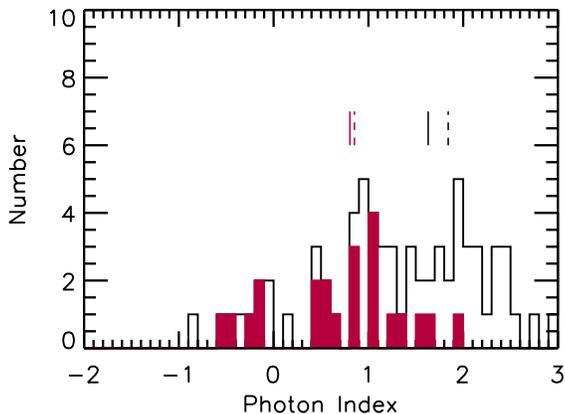}}
\caption{Effective observed photon index distribution for the central region
X-ray sample with $>3\sigma$ detections at 850~$\mu$m split by luminosity:
$L^{R}_{8-28~{\rm keV}} >10^{42.5}$~erg~s$^{-1}$ (red shaded histogram)
and lower X-ray luminosities (open histogram). The $\Gamma$ are from L17.
The means (solid lines) and medians (dashed lines) for the two samples 
are marked (red for the red shaded histogram, and black for the open histogram).
\label{psi_hist_smm}
}
\end{figure}

The bulk of the X-ray sources with speczs that are submillimeter detected 
(enclosed in open squares) are classified as SFGs (24) based on
their optical/NIR spectra. 
Most of these have a soft $\Gamma$ and a low X-ray luminosity,
consistent with being star formation dominated
(e.g., Sazonov \& Khabibullin 2017 give 
$\Gamma=2.1\pm0.1$ for the collective X-ray spectrum of luminous, 
high-mass X-ray binaries; black line in Figure~\ref{gamma}).
All six of the submillimeter detected SFGs with $\Gamma<0.7$ are 
at relatively low redshifts ($z<1.4$). These probably contain 
an obscured AGN.

Five Abs sources are also detected in the submillimeter, 
all at low redshifts ($z<1.1$). Four have soft $\Gamma$
consistent with being X-ray binary dominated, but one 
has a hard $\Gamma$.

However, before drawing conclusions based on the above results, 
we need to consider the possibility 
that we may be better able to measure speczs
in galaxies that are relatively unobscured and hence biased against 
submillimeter detections. To check this, we plot $\Gamma$ versus photz
for the central region X-ray sample with only photzs in Figure~\ref{gamma}(b).
There are 37 $>3\sigma$ submillimeter sources out of
347 in the specz sample of Figure~\ref{gamma}(a) ($10.7\pm1.7$\%)
and 38 out of 160 in the photz sample of Figure~\ref{gamma}(b) ($24\pm4$\%),
suggesting that submillimeter sources are indeed preferentially
missed in the specz sample (see also Figure~\ref{fs_kmg_redshifts}).

Since we do not have spectral classifications for the X-ray sources with only photzs,
we can only use $\Gamma$ and the estimated X-ray luminosity as diagnostics.
However, we can see from Figure~\ref{gamma}(b) that the population of
submillimeter detected X-ray sources (enclosed in open squares)
separates into a lower X-ray luminosity sample (smaller symbols)
with mostly soft $\Gamma$, consistent with these 
sources being star formers, and a more
X-ray luminous population (larger symbols) with mostly hard $\Gamma$.
These latter sources correspond to the intermediate X-ray luminosity
AGNs with submillimeter detections that we discussed in Section~\ref{subsecAGNSF}.

We can see this more clearly in Figure~\ref{psi_hist_smm}, where we show
the $\Gamma$ distribution for the submillimeter detected
X-ray sources divided into sources with $L^{R}_{8-28~{\rm keV}} >
10^{42.5}$~erg~s$^{-1}$ (red shaded histogram) and those with lower X-ray 
luminosities (open histogram). While there is clearly a large spread in $\Gamma$
for both X-ray luminosity samples, we find that the mean for
the red shaded histogram sample ($\bar{\Gamma}=0.55$) 
is significantly lower than 
the mean for the open histogram sample ($\bar{\Gamma}=1.58$).
These means are marked on the figure (solid lines), along with the medians
(dashed lines).
A Mann-Whitney test gives a one-sided probability of $10^{-5}$
that the two samples are drawn from the same distribution.
As expected, this suggests that most of the red shaded histogram sources
are dominated by highly absorbed AGN activity, while most of the open 
histogram sources are dominated by star formation activity; however, 
the latter sample also contains a small number of absorbed AGNs.

\begin{figure}[t]
\centerline{\includegraphics[width=9cm,angle=0]{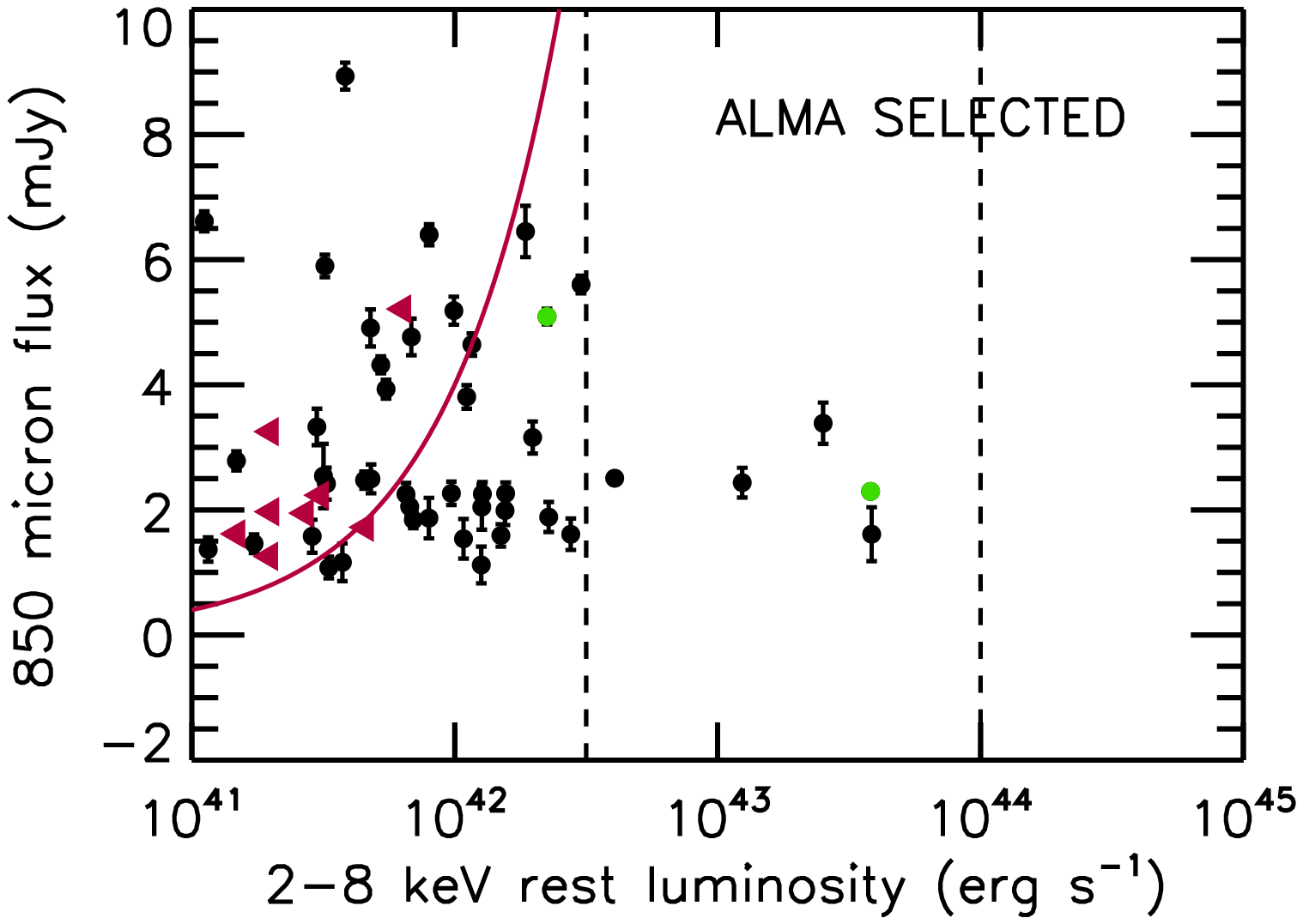}}
\centerline{\includegraphics[width=9cm,angle=0]{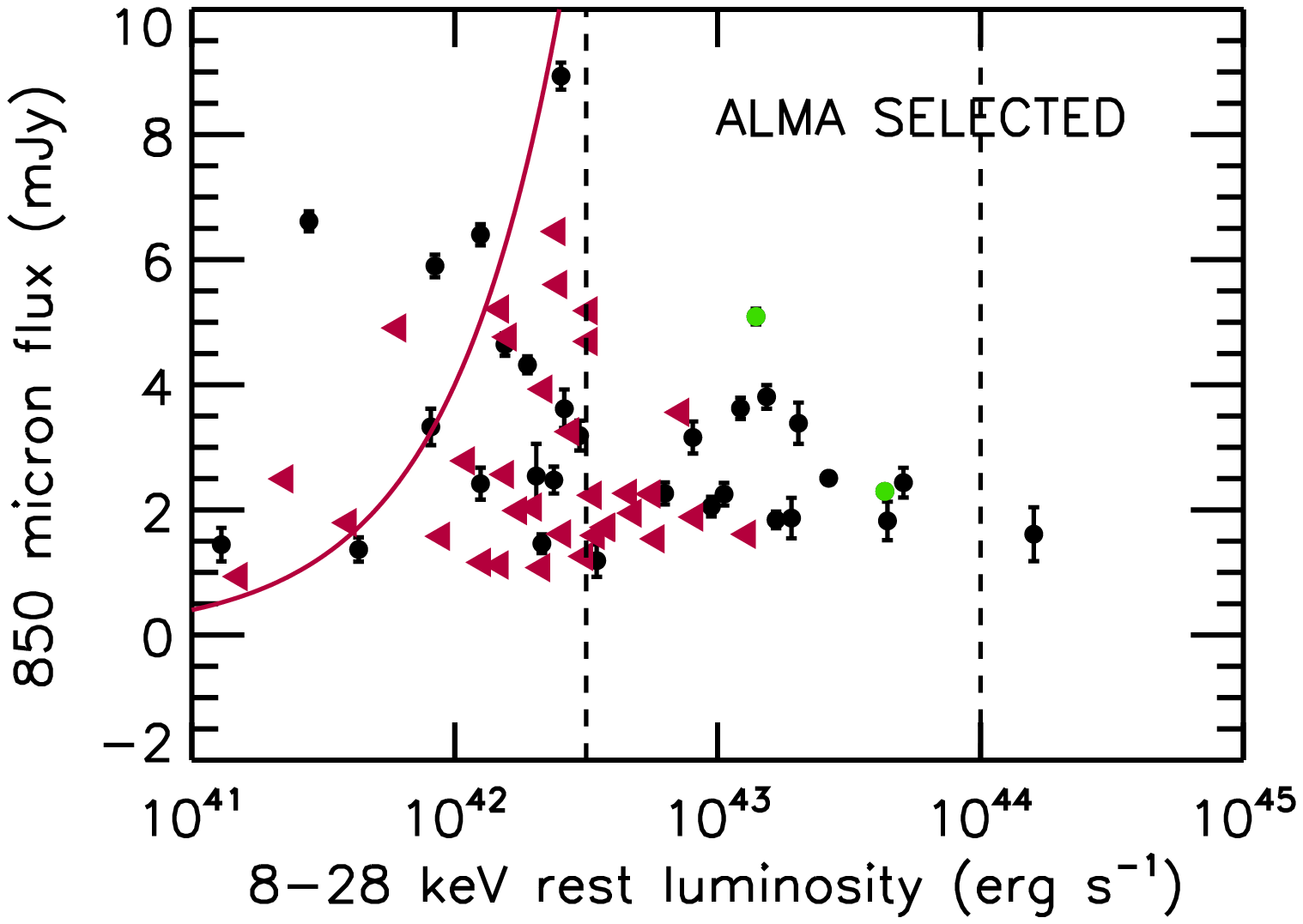}}
\caption{850~$\mu$m flux vs. (a) $L^{R}_{2-8~{\rm keV}}$
and (b) $L^{R}_{8-28~{\rm keV}}$ for the full ALMA sample
(black circles). In both panels, green circles show Sy2s.
No BLAGNs nor the BALQSO are present.
Sources not detected at the $>2\sigma$ level
in (a) 0.5--2~keV and (b) 2--7~keV are shown with red left-pointing 
triangles at the $2\sigma$ X-ray luminosity in that band. 
For sources with measured X-ray luminosities, we show
$\pm1\sigma$ uncertainties in the submillimeter flux.
The red curve shows the relation for sources where the 
X-rays are due to star formation. The dashed vertical lines
show $10^{42.5}$ and $10^{44}$~erg~s$^{-1}$.
\label{lx_smm_alma}
}
\end{figure}

\section{X-ray Properties of the ALMA Sample}
\label{almasec}
We now turn to the X-ray properties of the ALMA sample. 
The ALMA sample allows us to probe more deeply in X-rays
and to determine the fraction of submillimeter sources that
do not have X-ray counterparts, even at the depths of the CDF-S
7~Ms image.

In Figure~\ref{lx_smm_alma}, we plot 850~$\mu$m flux versus  
(a) $L^R_{\rm 2-8~keV}$ and
(b) $L^R_{\rm 8-28~keV}$ for the full ALMA sample.
The red curve shows the relation for sources where the X-rays are 
due to star formation (see Section~\ref{subsecAGNSF}). 
Clearly, the submillimeter picks out both star formers 
and intermediate X-ray luminosity AGNs in the X-rays, but not the most
luminous X-ray sources, which include the BLAGNs and many of the Sy2s.

\begin{figure*}[ht]
\centerline{\includegraphics[width=9cm,angle=0]{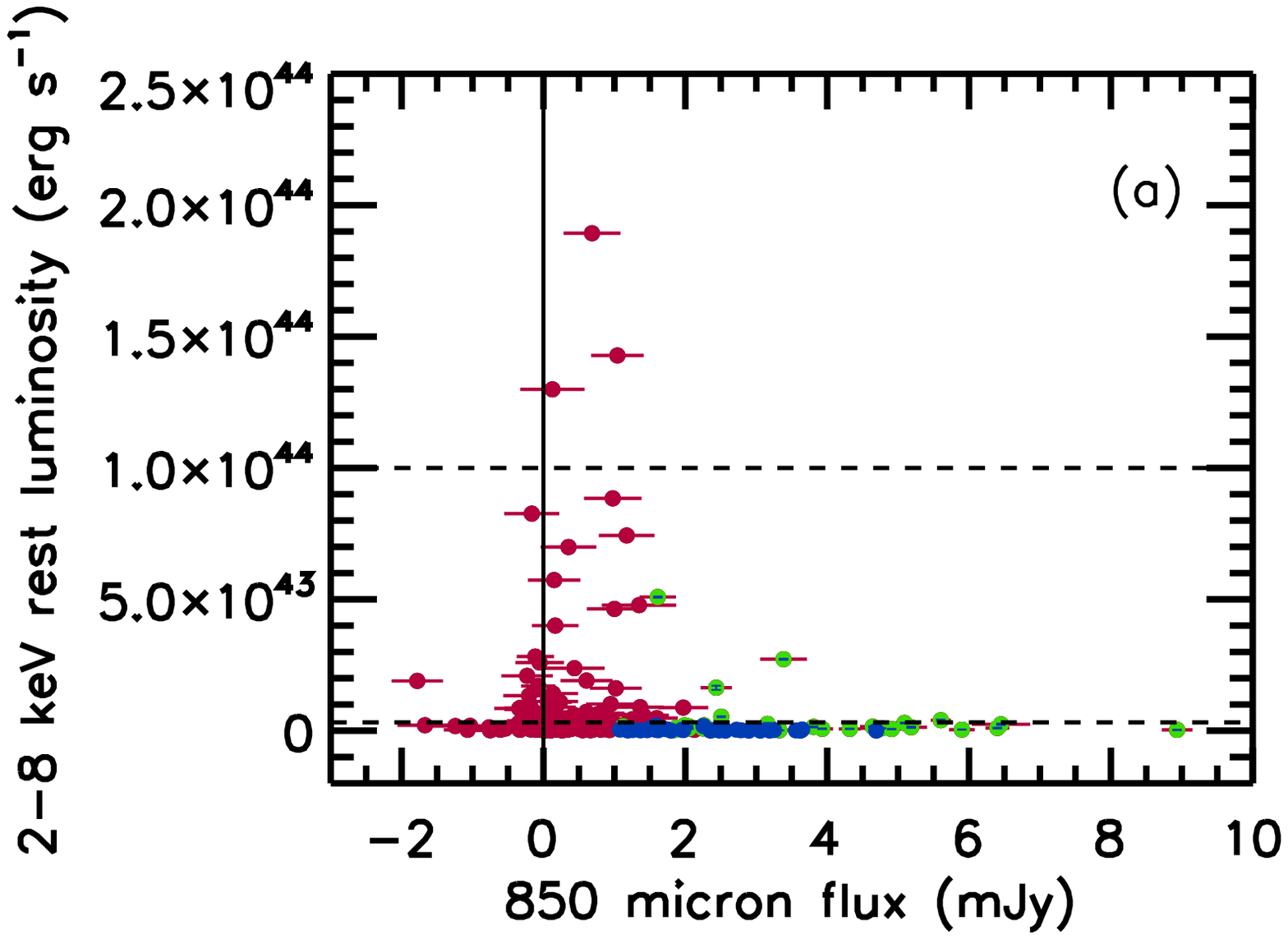}
\includegraphics[width=9cm,angle=0]{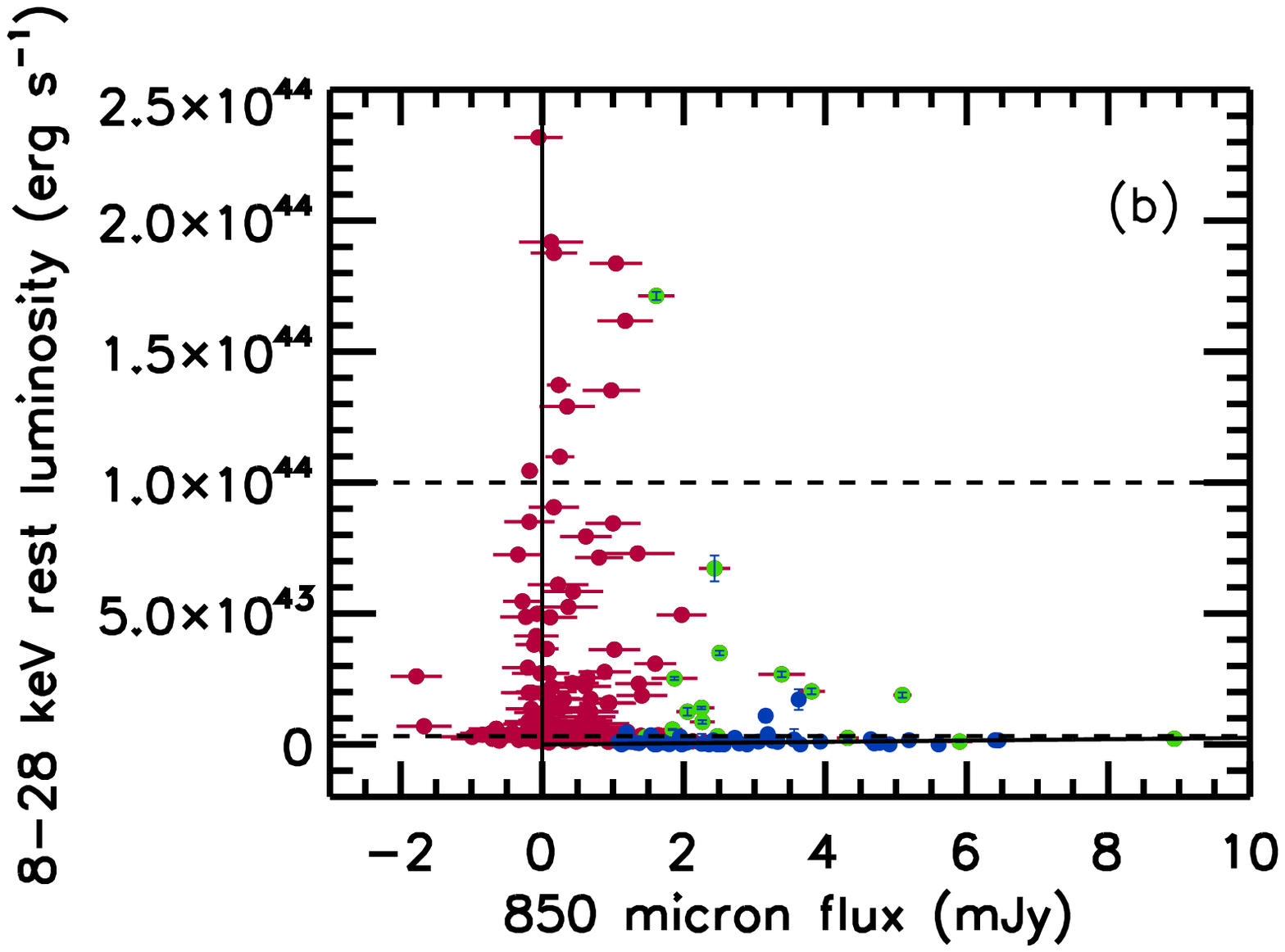}}
\centerline{\includegraphics[width=9cm,angle=0]{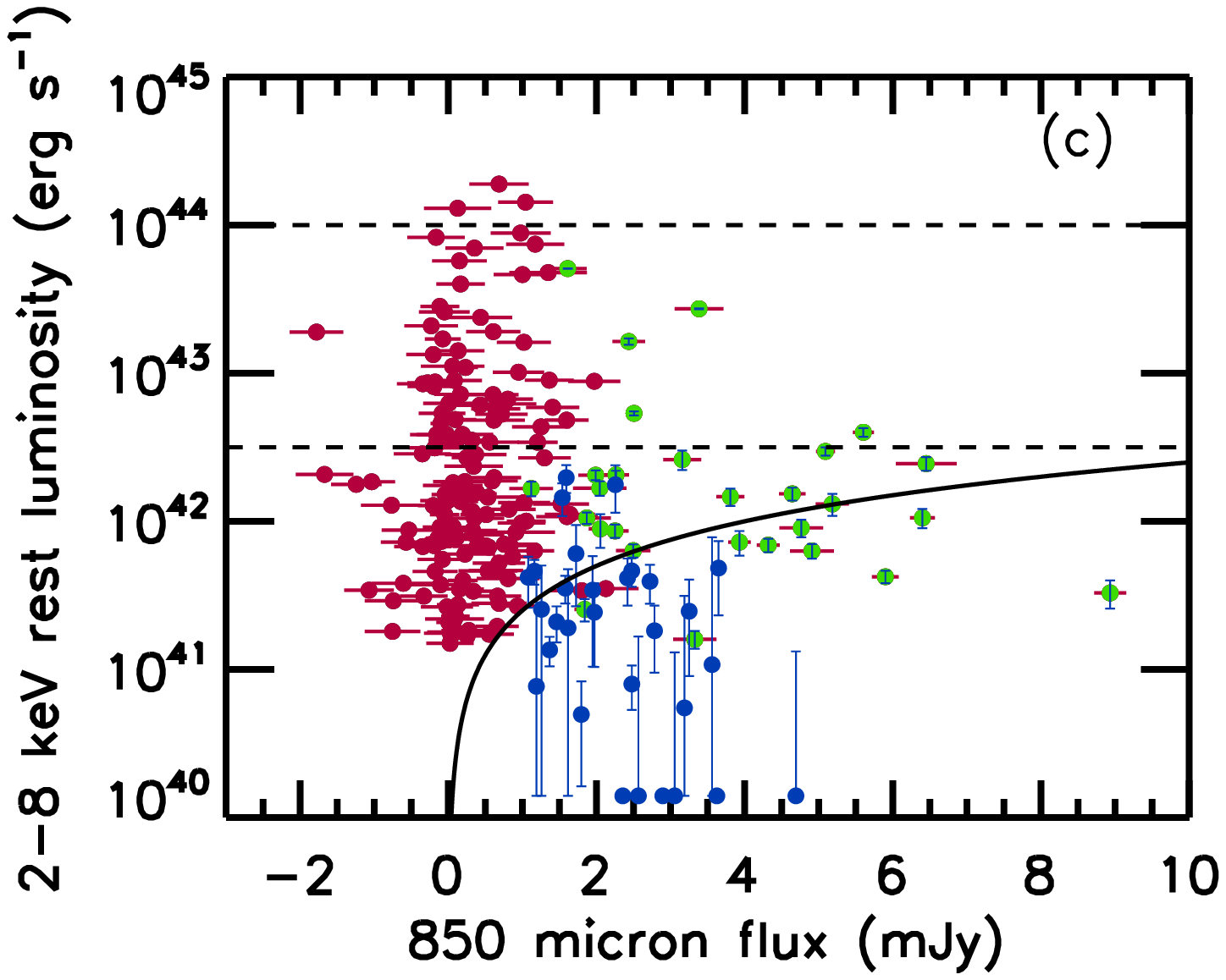}
\includegraphics[width=9cm,angle=0]{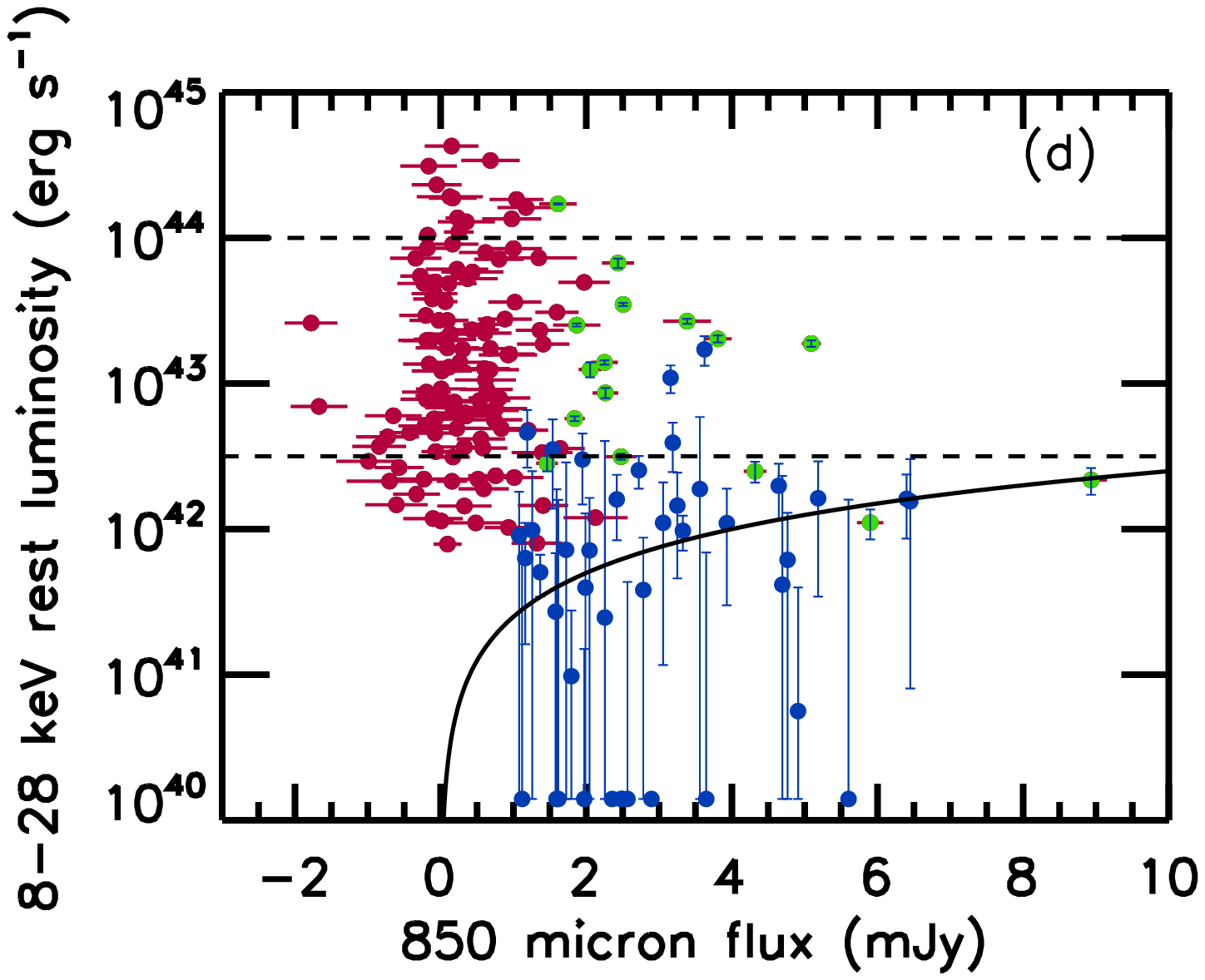}}
\caption{(a), (b) Linear and (c), (d) logarithmic representations
of (a), (c) $L^{R}_{\rm 2-8~keV}$ and (b), (d) 
$L^{R}_{\rm 8-28~keV}$ vs. 850~$\mu$m flux in the
central region, restricted to sources with 
$1<z<4.5$; the upper cut-off is to avoid the highly uncertain photzs of
high-redshift candidates (see Cowie et al.\ 2019).
$L^{R}_{\rm 2-8~keV}$ or $L^{R}_{\rm 8-28~keV}<10^{40}$~erg~s$^{-1}$ 
sources are placed at $\sim10^{40}$~erg~s$^{-1}$. 
Red circles show sources from the X-ray sample
((a), (c), restricted to 
0.5--2~keV fluxes above $2\times 10^{-17}$~erg~cm$^{-2}$~s$^{-1}$,
and (b), (d), restricted to
2--7~keV fluxes above $10^{-16}$~erg~cm$^{-2}$~s$^{-1}$),
blue from the ALMA sample (note that the X-ray luminosities are 
based on the redshifts given in Table~2, and photzs 
for submillimeter sources can be quite uncertain; see C18),
and green are present in both. We show $\pm1\sigma$ uncertainties in the 
850~$\mu$m flux for the X-ray sources, and $\pm1\sigma$ uncertainties in
the X-ray luminosity for the submillimeter sources.
In (c) and (d), the black curve shows the relation for sources where the X-rays
are due to star formation.
The dashed horizontal lines
show $10^{42.5}$ and $10^{44}$~erg~s$^{-1}$.
\label{compare_smm_xray}
}
\end{figure*}

More specifically, we can see that in Figure~\ref{lx_smm_alma}(a), all but 
five of the ALMA sources
have low X-ray luminosities ($<10^{42.5}$~~erg~s$^{-1}$).
Intermediate X-ray luminosity AGNs, because of their high optical depths, 
are more easily seen in Figure~\ref{lx_smm_alma}(b).
However, even for this 2--7~keV band, which has
much poorer sensitivity, most of the sources are consistent with being
star formation dominated, at least in as far as the upper limits
are constraining. The ALMA sample picks out one high and 15 intermediate 
X-ray luminosity AGNs.

Finally, in Figure~\ref{compare_smm_xray},
we combine the two samples and plot (a), (c) $L^{R}_{2-8~{\rm keV}}$ and 
(b), (d) $L^{R}_{\rm 8-28~keV}$ versus submillimeter flux 
for both the central X-ray (red) and ALMA (blue) samples;
sources included in both samples are green. 
We show linear representations in the (a), (b) plots,
since the X-ray luminosities and submillimeter fluxes may be 
negative, and these plots are helpful
in visualizing the large number of low luminosity sources,
but we also show the logarithmic versions in the (c), (d)
plots to provide a clearer representation of the full range
of X-ray luminosities.
It is again evident that most of the ALMA sources 
(blue and green) are not very luminous in X-rays. 
For sources with 850~$\mu$m fluxes above 2.25~mJy, only 10 of the 36
sources have $L^{R}_{\rm 8-28~keV}>10^{42.5}$~erg~s$^{-1}$,
while above 4~mJy, only 1 of the 11 sources does. 

In Figures~\ref{compare_smm_xray}(c), (d),
the black curve shows the relation 
for the sources where the X-rays are due to star formation.
Most of the ALMA sources, both with (green) and without (blue) 
X-ray detections, appear to be consistent with having both their X-ray 
and their submillimeter emission driven by star formation, even for the harder
$L^{R}_{\rm 8-28~keV}$. However, a small population of 
ALMA sources with intermediate X-ray luminosities is only 
clearly seen in Figures~\ref{compare_smm_xray}(b), (d),
reflecting their high obscuration (see Section~\ref{secgamma}).

\section{Discussion}
\label{secdisc}
\begin{figure}
\centerline{\includegraphics[width=9cm,angle=0]{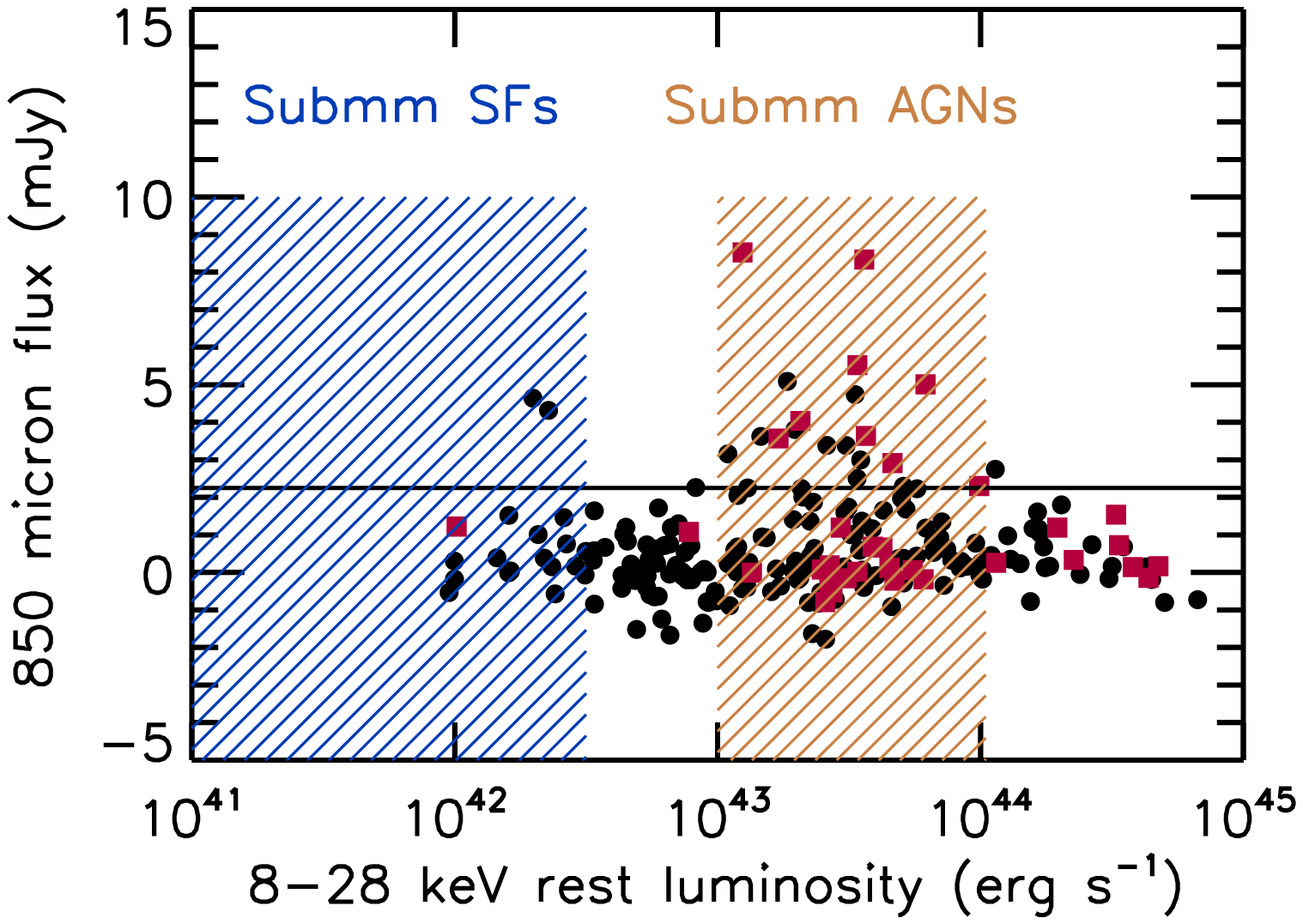}}
\caption{850~$\mu$m flux vs. $L^R_{\rm 8-28~keV}$ for
$z=1.5$--4.5 (specz and photz) sources
in the CDF-S (black circles) and CDF-N (red squares) lying at off-axis 
angles $<8'$, where the rms noise is $<1$~mJy for all sources.
The black horizontal line shows a submillimeter flux of 2.25~mJy (322~M$_\sun$~yr$^{-1}$),
which roughly corresponds to the SCUBA-2 $>4\sigma$ limit through the whole 
central region. We consider sources above this to be strong submillimeter sources.
The blue shaded region shows where strong submillimeter sources are found
for X-ray luminosities powered by star formation, and the gold shaded region
shows where strong submillimeter sources are found for X-ray luminosities 
powered by AGNs.
\label{lx_smm_plus_cdfn}
}
\end{figure}

In order to search for variance and to increase the
size of the sample for statistical analyses, we now combine the 
CDF-S with the CDF-N, where we have the 2~Ms X-ray image from 
Alexander et al.\ (2003) and the submillimeter data (SCUBA-2 and
SMA) from C17.
In Figure~\ref{lx_smm_plus_cdfn}, we show submillimeter 
flux versus $L^{R}_{8-28~{\rm keV}}$ for $z=1.5$--4.5 (specz and photz)
sources in the CDF-S (black circles) and CDF-N (red squares)
lying at off-axis angles $<8'$, where the rms noise is $<1$~mJy 
for all sources. We will hereafter refer to this as the 
{\em combined X-ray sample\/}.

Both fields have a number of strong submillimeter sources
(850~$\mu$m fluxes $>2.25$~mJy, black line)
at intermediate X-ray luminosities powered by AGNs
(gold shaded region; hereafter, we refer to these sources as submillimeter AGNs), 
but not at high X-ray luminosities. Indeed, the distributions of the submillimeter fluxes
are significantly different: a Mann-Whitney 
test gives only a 4\% probability that sources just below the highest 
X-ray luminosities
(namely, $L^{R}_{8-28~{\rm keV}}=10^{43.5}$--$10^{44}$~erg~s$^{-1}$) 
are drawn from the same distribution as sources at
$L^{R}_{8-28~{\rm keV}}>10^{44}$~erg~s$^{-1}$.

We note that Page et al.\ (2012) found a similar result using
low resolution and much shorter wavelength {\em Herschel\/}-SPIRE 250~$\mu$m data on 
only the CDF-N. Based on a combined {\em Herschel\/}-SPIRE 250~$\mu$m data sample 
from the CDF-N, CDF-S, and COSMOS, Harrison et al.\ (2012) argued that the Page et al.\ 
result was likely a consequence of poor source statistics and potentially cosmic variance. 
We emphasize that ultimately much larger samples will be required to provide very robust results.

A Kolmogorov-Smirnov test shows that the submillimeter fluxes
for X-ray sources in the range 
$L^{R}_{8-28~{\rm keV}}=10^{43}$--$10^{44}$~erg~s$^{-1}$
have a $<1$\% probability of being normally distributed.
They have a mean flux of 1.00~mJy 
with a 95\% confidence range 0.59--1.42~mJy. In contrast,
the submillimeter fluxes for X-ray sources at
$L^{R}_{8-28~{\rm keV}}>10^{44}$~erg~s$^{-1}$ are consistent
with being normally distributed (probability of 0.31).
They have a mean flux of
0.58~mJy with a 95\% confidence range 0.19--0.98~mJy.
This corresponds to a drop in the mean SFR from the
intermediate to the high X-ray luminosity range of
$\sim140$~M$_\odot$~yr$^{-1}$ to $\sim80$~M$_\odot$~yr$^{-1}$.

\begin{figure}
\centerline{\includegraphics[width=9cm,angle=0]{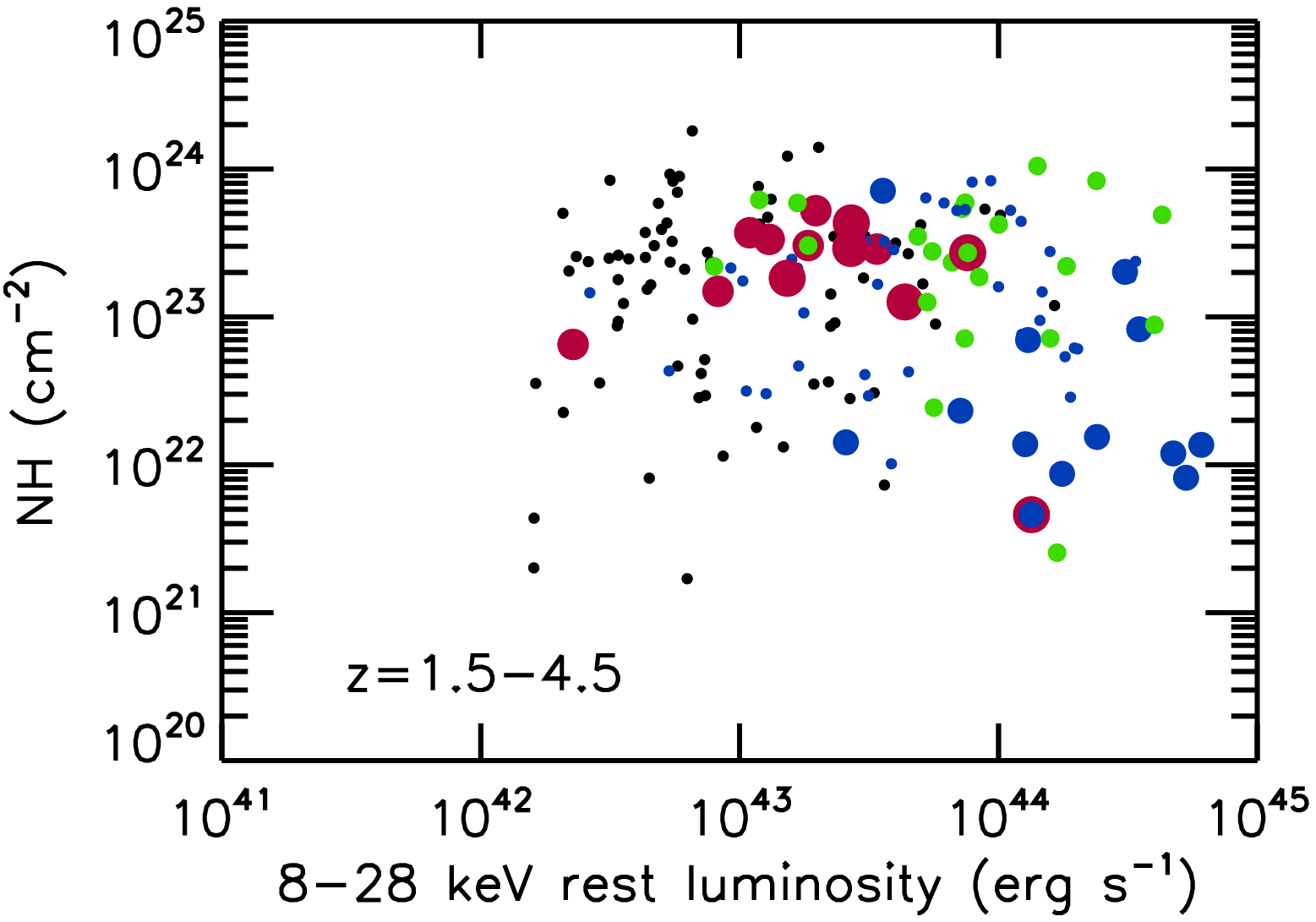}}
\centerline{\includegraphics[width=9cm,angle=0]{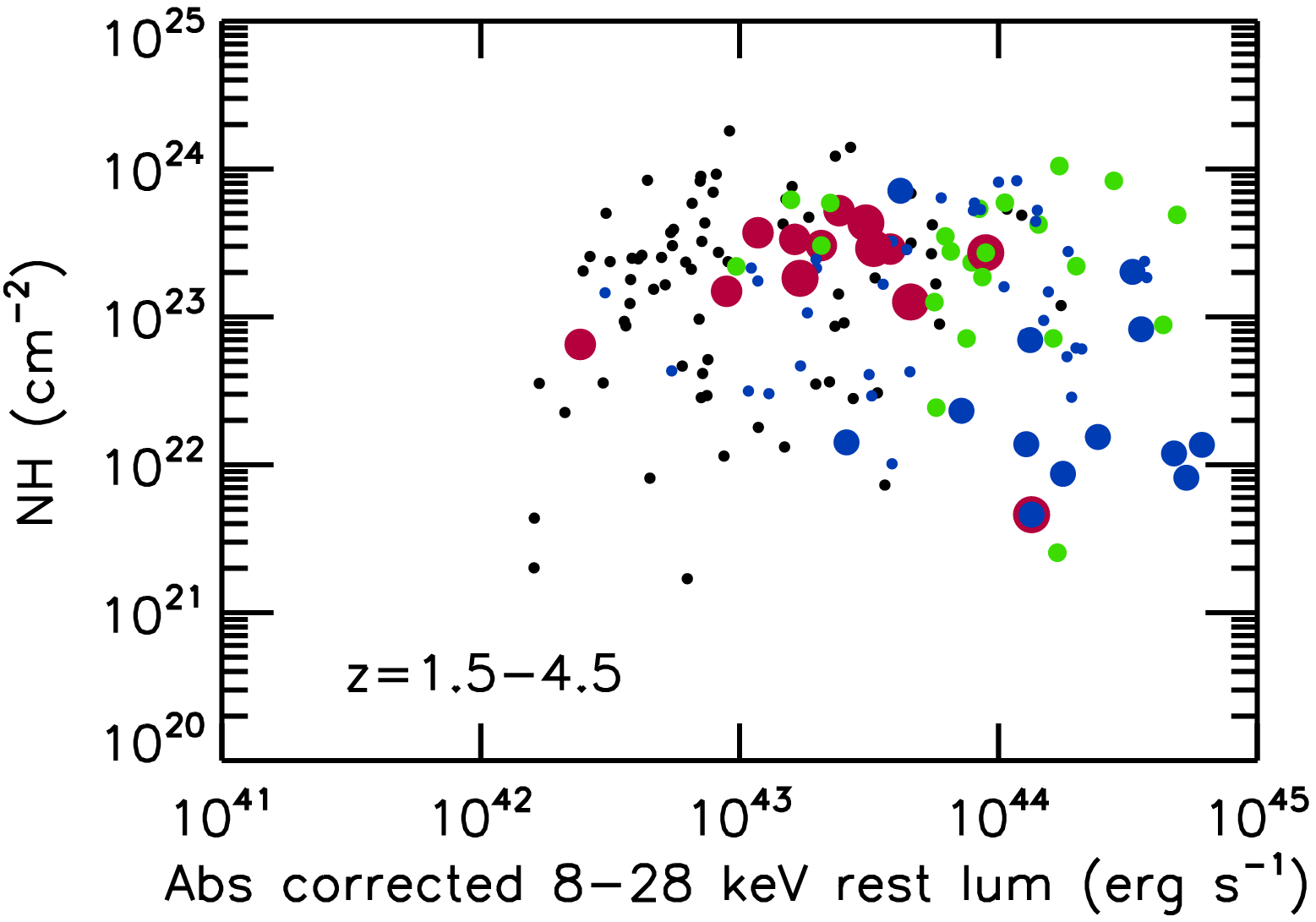}}
\centerline{\includegraphics[width=9cm,angle=0]{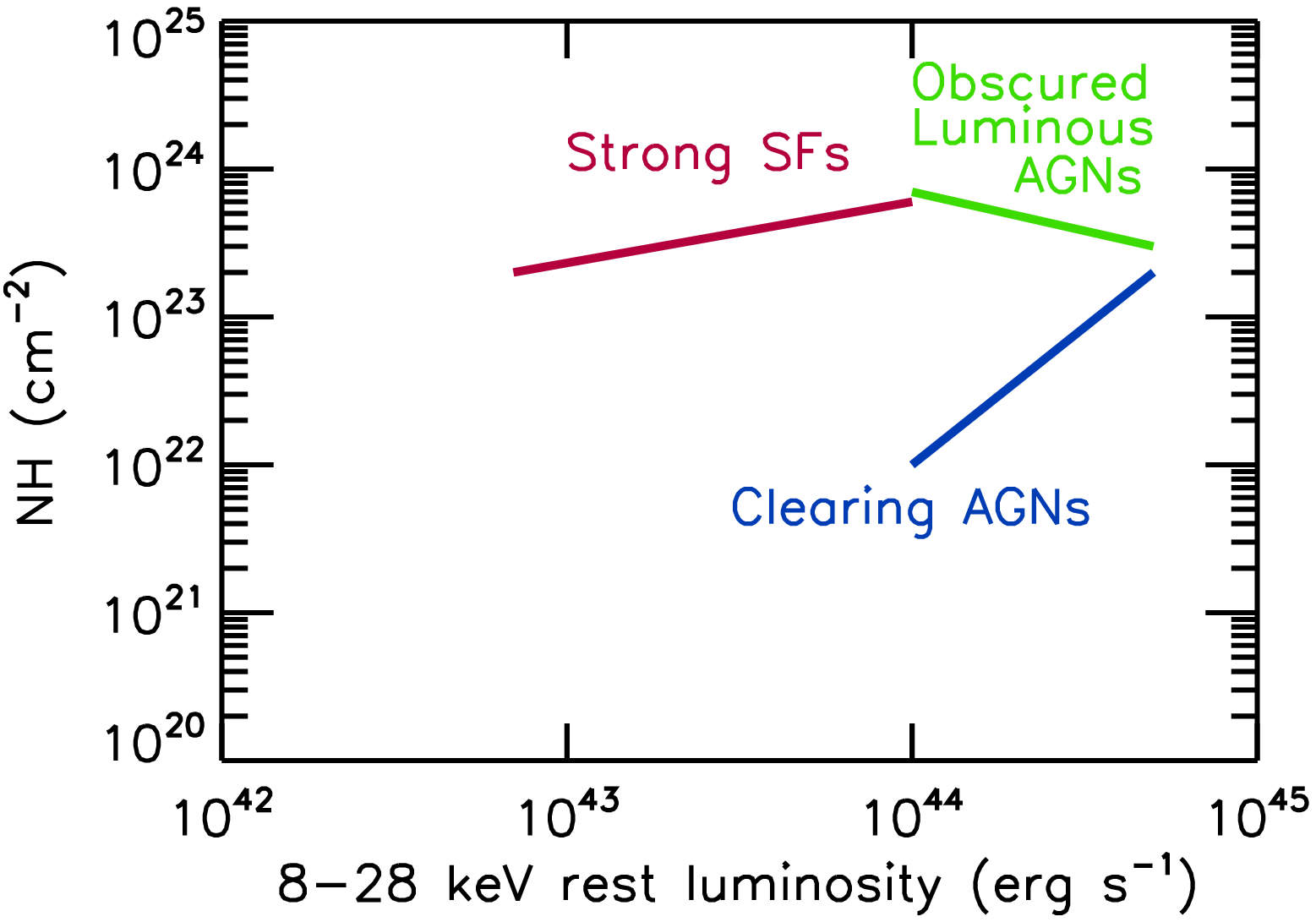}}
\caption{X-ray absorption column density, $N_H$, vs. 
(a) $L^{R}_{8-28~{\rm keV}}$ and (b) absorption-corrected
$L^{R}_{8-28~{\rm keV}}$ for
$z=1.5$--4.5 (specz and photz) sources
in the CDF-S and CDF-N lying at off-axis 
angles $<8'$, where the rms noise is $<1$~mJy for all sources.
Strong submillimeter sources ($>2.25$~mJy,
or $>322~{\rm M}_\odot$~yr$^{-1}$) are shown with red circles.
Spectrally classified BLAGNs are shown in blue, and Sy2s in green. 
Sources that are both a strong submillimeter source and have a 
BLAGN or Sy2 spectral classification are shown as red circles
with blue or green interiors, respectively. The only BLAGN in
this category comes from the CDF-N.
(c) Schematic outline with the location of the various stages labeled.
\label{lx_nh}
}
\end{figure}

As discussed in Section~\ref{secgamma}, the submillimeter AGNs
correspond to sources with high absorption in their X-ray spectra. 
In Figure~\ref{lx_nh}(a), we show X-ray absorption column density,
$N_H$, versus $L^{R}_{8-28~{\rm keV}}$
for the combined X-ray sample.
We calculate $N_H$ by assuming the intrinsic power law is 1.8 and then
comparing with the individual effective observed $\Gamma$ values from L17.
We show the strong submillimeter sources
with large red circles. We also show sources spectrally classified as 
BLAGNs (blue circles) or Sy2s (green circles). There are a small number 
of submm AGNs that have a 
BLAGN or Sy2 spectral classification, which we denote by
red circles with blue or green interiors, respectively.
(Note that the only BLAGN in this category comes from the CDF-N.)

As we discussed in Section~\ref{subsecxraylum} and
stressed in Section~\ref{subsecAGNSF},
we have not corrected the X-ray luminosities for any intrinsic absorption.
For high redshifts, the 2--7~keV band corresponds to very high 
energies, so these corrections are small. 
We use $N_H$ to calculate the absorption corrections in each of the energy bands.
We show $N_H$ versus absorption-corrected $L^{R}_{8-28~{\rm keV}}$
in Figure~\ref{lx_nh}(b). We can see the effect is very minor, moving the
high $N_H$ sources to slightly higher luminosities relative to the lower 
$N_H$ sources. 

We see that the submm AGNs primarily
have high $N_H$, while the Sy2s lie, on average, at higher X-ray 
luminosities but still have high $N_H$, and the BLAGNs lie at the highest
X-ray luminosities but have lower $N_H$. 
We schematically illustrate this in Figure~\ref{lx_nh}(c).
 
It is plausible to interpret this result in terms of the Sanders-Hopkins
merger model (Sanders et al.\ 1988; Hopkins et al.\ 2006).
After the early stages of the 
merger process, an intense star formation period occurs (hundreds of
solar masses per year), which we would identify with
the strong submillimeter sources. This is followed by declining star formation
and high X-ray luminosity, which we would identify with the Sy2 population.
Then, as the gas is cleared from the system but the X-ray
luminosity remains high, we would identify the source as a BLAGN. 
The timescales of each episode are comparable in this interpretation
so that we would expect to see similar numbers of sources
in each stage, as is indeed the case.

\section{Summary}
\label{secsummary}
We used the X-ray catalog of L17 for the CDF-S to construct a full X-ray sample of 938 
sources (off-axis angles $<10'$) complemented with SCUBA-2 850~$\mu$m 
data with noise $<1.5$~mJy (C18). We also constructed a central region 
X-ray sample of 526 sources (off-axis angles $<5\farcm7$) 
complemented with SCUBA-2 850~$\mu$m data with noise 
$<0.56$~mJy (this corresponds to a $4\sigma$ limit
of $>2.25$~mJy) and deep ALMA band~7 data (75 sources at $>4.5\sigma$; C18). 

We assigned secure speczs
to 64.5\% of the sources in the full X-ray sample and 
spectral classifications to 64\%. However, for the $>3\sigma$ submillimeter 
detected X-ray sources, the percentage of speczs
is much smaller (50\%), likely due to the fact that these sources are dustier, 
fainter in the optical/NIR, and at higher redshifts.
After including photzs from the literature (H14 and S16), all but 44 
of the sources in the full X-ray sample have a redshift. 
Most of the X-ray sources lie at low redshifts (58\% at $z<1.6$,
88\% of which are speczs), while only a handful lie at high redshifts 
(9 sources at $z>4$ and 2 sources at $z>5$, nearly all of which are photzs;
see Cowie et al.\ 2019 for
a detailed discussion of these very high-redshift AGN candidates).

We presented a catalog of the
X-ray and submillimeter properties, redshifts, and spectral classes of the full X-ray sample.
We also presented a catalog of the X-ray properties and redshifts of the 75 ALMA sources.

Our major results are as follows:

\begin{itemize}

\item[$\bullet$]
Most of the spectrally classified X-ray sources with $>3\sigma$ 850~$\mu$m
detections are star-forming galaxies; very few Sy2s and no BLAGNs in 
either the full or central X-ray samples have 850~$\mu$m detections at $>3\sigma$.

\item[$\bullet$] 
The X-ray sources contain a significant 850~$\mu$m flux with 
a mean of 0.52~mJy per X-ray source with a 95\% confidence range
of 0.44 to 0.61~mJy. However, nearly all of the signal comes from those
X-ray sources that are significantly detected in the submillimeter. 
The bulk of the X-ray sources are consistent with having no 850~$\mu$m flux.
This emphasizes the dangers of using simple stacking 
analyses on samples where the sources producing most of the 
850~$\mu$m flux lie in a very extended tail.

\item[$\bullet$]
Only $\sim10$\% of the 850~$\mu$m EBL measured by Fixsen et al.\ (1998)
is identified by X-ray sources (sources where the X-ray emission is powered
by AGNs and sources where the X-ray emission is powered by star formation) 
at the limits of the {\em Chandra\/} 7~Ms data. Thus, Compton-thin X-ray detected
AGNs, which we are sensitive to out to $z\sim6$, do not mark 
the galaxies responsible for producing most of the 850~$\mu$m light.

\item[$\bullet$]
Our analysis uses X-ray luminosities uncorrected for absorption (except in the Discussion),
but for high redshifts, the 2--7~keV band corresponds to very high energies, so
these corrections are small.
The brightest 850~$\mu$m detected X-ray sources are primarily at low X-ray
luminosities ($L^{R}_{8-28~{\rm keV}} <10^{42.5}$~erg~s$^{-1}$). In fact, all 
but one of the central region X-ray sources with an
850~$\mu$m flux $>4$~mJy are consistent with both the X-ray luminosity and
the 850~$\mu$m flux being produced by star formation.

\item[$\bullet$]
At X-ray luminosities that can only be produced by an AGN
($L^{R}_{8-28~{\rm keV}} > 10^{42.5}$~erg~s$^{-1}$), only 8 sources are
strong submillimeter sources (850~$\mu$m flux $>2.25$~mJy), all of which 
are at intermediate X-ray luminosities
($10^{42.5} < L^{R}_{8-28~{\rm keV}} < 10^{44}$~erg~s$^{-1}$). 
Thus, it appears that extreme SFRs\,$\gtrsim 300$~M$_\odot$~yr$^{-1}$ are 
only seen in the hosts of intermediate X-ray luminosity AGNs. 

\item[$\bullet$]
Consistent with the above results, the distribution of the effective observed
photon indices for the 850~$\mu$m detected
$L^{R}_{8-28~{\rm keV}} > 10^{42.5}$~erg~s$^{-1}$ sources ($\bar{\Gamma}=0.55$) 
is significantly different than the distribution for the 850~$\mu$m detected
lower X-ray luminosity sources ($\bar{\Gamma}=1.58$); this suggests that the
former are dominated by highly absorbed AGN activity, 
while the latter are dominated by star formation activity.

\item[$\bullet$]
Starting with the ALMA sample makes it possible to probe the X-ray data more
deeply.  Most of the ALMA sources appear to be consistent 
with having both their X-ray and their 850~$\mu$m emission driven by 
star formation, with only a small population having intermediate X-ray luminosities.

\item[$\bullet$]
Our results are consistent with seeing the effects of gas being cleared and 
the SFR being lowered as the galaxy becomes more 
transparent and X-ray luminous.

\end{itemize}

\vskip 2.5cm
\acknowledgements
We thank the anonymous referee for a careful report that helped us to improve the manuscript.
We gratefully acknowledge support from NSF grant AST-1313150 (A.~J.~B.),
NASA grant NNX17AF45G (L.~L.~C.),
CONICYT grants Basal-CATA PFB-06/2007 (F.~E.~B, J.~G.-L.) and
Programa de Astronomia FONDO ALMA 2016 31160033 (J.~G.-L.),
and the Ministry of Economy, Development, and Tourism's Millennium Science 
Initiative through grant IC120009, awarded to The Millennium Institute of Astrophysics, 
MAS (F.~E.~B.).
Este trabajo cont{\'o} con el apoyo de CONICYT + Programa de Astronom{\'i}a + 
Fondo CHINA-CONICYT CAS16026 (J.~G.-L.).
Support for this research was also provided by the University of Wisconsin-Madison,
Office of the Vice Chancellor for Research and Graduate Education with funding
from the Wisconsin Alumni Research Foundation,
the John Simon Memorial Guggenheim Foundation, and the Trustees
of the William F. Vilas Estate (A.~J.~B.). 
A.~J.~B would like to thank the Lorentz Center for
the stimulating workshop ``Monsters of the Universe:  The Most Extreme Star Factories''
that benefited this work.
ALMA is a partnership of ESO (representing its member states), 
NSF (USA) and NINS (Japan), together with NRC (Canada), 
MOST and ASIAA (Taiwan), and KASI (Republic of Korea), in cooperation 
with the Republic of Chile. The Joint ALMA Observatory is operated by 
ESO, AUI/NRAO and NAOJ.
The James Clerk Maxwell Telescope is operated by the East Asian Observatory 
on behalf of The National Astronomical Observatory of Japan, Academia Sinica 
Institute of Astronomy and Astrophysics, the Korea Astronomy and Space 
Science Institute, the National Astronomical Observatories of China and the 
Chinese Academy of Sciences (Grant No. XDB09000000), with additional funding 
support from the Science and Technology Facilities Council of the United Kingdom 
and participating universities in the United Kingdom and Canada.
The W.~M.~Keck Observatory is operated as a scientific
partnership among the California Institute of Technology, the University
of California, and NASA, and was made possible by the generous financial
support of the W.~M.~Keck Foundation.
The authors wish to recognize and acknowledge the very significant 
cultural role and reverence that the summit of Maunakea has always 
had within the indigenous Hawaiian community. We are most fortunate 
to have the opportunity to conduct observations from this mountain.


\clearpage
\startlongtable
\begin{deluxetable}{cccccccccccc}
\renewcommand\baselinestretch{1.0}
\tablewidth{0pt}
\tablecaption{X-ray and Submillimeter Properties, Redshifts, and Spectral Classes of the Full X-ray Sample}
\scriptsize
\tablehead{L17 & R.A. & Decl.& $f_{\rm 0.5-2~keV}$ & $f_{\rm 2-7~keV}$ & $f_{\rm 850~\mu m}$ & Error & ALMA & $z_{spec}$ & Spectral & Reference  & $z_{phot}$ \\ 
No. & J2000.0 & J2000.0 
& \multicolumn{2}{c}{(10$^{-17}$~erg~cm$^{-2}$~s$^{-1}$)} & \multicolumn{2}{c}{(mJy)} & Flag & & Class & \\ 
(1) & (2) & (3) & (4) & (5) & (6) & (7) & (8) & (9) & (10) & (11) & (12) }
\startdata
518 &   53.118004 &  -27.802055 &   1.96 &  -1.07 & -0.45 & 0.25 & 0 & 0.000 & Star & aqp  & 0.446\cr
487 &   53.111496 &  -27.802973 &   1.21 &  -3.24 & 0.068 & 0.25 & 0 & \nodata & \nodata &   & 0.732\cr
530 &   53.120167 &  -27.798832 &   2.20 &  -2.88 &   1.0 & 0.26 & 0 & \nodata & \nodata &   &  2.340\cr
482 &   53.110790 &  -27.800556 &   1.18 &  -4.90 & .0086 & 0.15 & 1 & \nodata & \nodata &   &  2.731\cr
491 &   53.111706 &  -27.813999 &   1.25 &  -2.64 & 0.021 & 0.25 & 0 & \nodata & \nodata &   &  2.585\cr
470 &   53.108250 &  -27.797583 &   3.17 &  -2.29 & 0.065 & 0.17 & 1 & \nodata & \nodata &   &  1.895\cr
575 &   53.128834 &  -27.813749 & -0.586 &   6.84 &  0.56 & 0.26 & 0 & \nodata & \nodata &   &  2.394\cr
442 &   53.103870 &  -27.804028 &   1.33 &  -2.65 & -0.36 & 0.21 & 0 & 0.128 & Abs & q  & 0.144\cr
578 &   53.129749 &  -27.799028 &  0.774 &  -4.88 &  0.36 & 0.26 & 0 & \nodata & \nodata &   &  2.731\cr
531 &   53.120750 &  -27.818972 &   1.68 &  -8.06 &   1.4 & 0.26 & 0 &  1.094 & Abs & abh  &  1.075\cr
589 &   53.131332 &  -27.814917 &   11.5 &   16.1 &   1.2 & 0.27 & 0 & \nodata & \nodata &   &  1.917\cr
432 &   53.102543 &  -27.814278 &   1.47 &  -4.99 &  0.75 & 0.25 & 0 & 0.577 & SFR & adp  & \nodata\cr
598 &   53.134293 &  -27.812611 &   1.67 &  -4.10 &  0.20 & 0.27 & 0 & 0.535 & SFR & aqph  & 0.525\cr
478 &   53.110081 &  -27.791750 & -0.892 &   5.35 & 0.035 & 0.26 & 0 & \nodata & \nodata &   & \nodata\cr
426 &   53.099918 &  -27.808416 &   1.06 &   7.44 &  0.38 & 0.25 & 0 & \nodata & \nodata &   &  1.969\cr
431 &   53.102543 &  -27.814278 &   2.41 &   3.47 &  0.70 & 0.25 & 0 & 0.579 & SFR & adp  & 0.484\cr
477 &   53.109539 &  -27.820805 &   2.42 &   7.49 &  0.30 & 0.25 & 0 & 0.338 & SFR & ah  & 0.331\cr
533 &   53.120834 &  -27.823055 &  -1.23 &   13.2 &  0.89 & 0.26 & 0 & 0.735 & Abs & bp  & 0.741\cr
616 &   53.137749 &  -27.802084 &   2.69 &   9.00 &  0.13 & 0.28 & 0 &  1.188 & Abs & qh  &  1.202\cr
580 &   53.130417 &  -27.791054 &   5.37 &   29.9 & 0.093 & 0.28 & 0 & \nodata & \nodata &   &  3.240\cr
481 &   53.110580 &  -27.823610 &   3.52 &  -2.26 &  0.86 & 0.26 & 0 &  1.468 & SFR & abh  &  1.415\cr
604 &   53.136333 &  -27.816528 &   1.25 &  -2.22 & 0.094 & 0.28 & 0 & 0.670 & SFR & dh  & 0.667\cr
583 &   53.130627 &  -27.790222 &   3.19 &  -3.32 &  0.11 & 0.28 & 0 & 0.666 & SFR & aqeh  & 0.677\cr
609 &   53.137165 &  -27.815638 &  0.884 &  -2.29 &  0.34 & 0.28 & 0 & 0.837 & SFR & ah  & 0.838\cr
417 &   53.097248 &  -27.814554 &   3.76 &   4.85 &  0.33 & 0.25 & 0 & 0.248 & SFR & adefh  & 0.248\cr
537 &   53.121712 &  -27.785418 &   2.51 &  -3.51 &  0.36 & 0.28 & 0 & 0.669 & Abs & abqef  & 0.676\cr
508 &   53.115498 &  -27.827055 & -0.956 &  -5.95 & -0.46 & 0.27 & 0 & 0.227 & SFR & ah  & \nodata\cr
401 &   53.094086 &  -27.804167 &   2.04 &   46.2 &   1.8 & 0.13 & 1 &  1.299 & SFR & b  &  2.391\cr
429 &   53.101501 &  -27.821556 &   1.42 &  -3.85 & -0.32 & 0.26 & 0 & \nodata & \nodata &   &  2.425\cr
\enddata
\label{tab2}
\tablecomments{
Excerpt of full table, which is available in its entirety in machine-readable form in the journal.
The table is ordered by {\em Chandra\/}
off-axis angle, such that the 526 sources in the central region X-ray sample
appear first, extending down to source L17~\#492.
The columns are (1) X-ray source number from L17, 
(2) and (3) optical/NIR R.A. and Decl.,
(4) and (5) observed 0.5--2~keV and 2--7~keV fluxes,
(6) and (7) observed 850~$\mu$m flux and error,
(8) a ``1" if the submillimeter flux measurement is based on ALMA data and
a ``0" if it is based on SCUBA-2 data,
(9) specz, (10) spectral class, (11) reference for the spectra used in the spectral classification
(a=our Keck DEIMOS, b=Vanzella et al.\ 2008, c,d=Balestra et al.\ 2010, 
e=Szokoly et al.\ 2004, f=Silverman et al.\ 2010, g=Kurk et al.\ 2013, 
h=Inami et al.\ 2017, Urrutia et al.\ 2019, i=McLure et al.\ 2018, j=our Keck MOSFIRE, 
k=Kriek et al.\ 2015, l=our Keck LRIS, m=Casey et al.\ 2011, n=Morris et al.\ 2015, 
o=Cooper et al.\ 2012, p=Le F{\`e}vre et al.\ 2005, q=Mignoli et al.\ 2005),
and (12) photz from H14, or, if there is no photz in H14, then from S16.
}
\end{deluxetable}
\onecolumngrid

\startlongtable
\begin{deluxetable}{ccccccccccccc}
\renewcommand\baselinestretch{1.0}
\tablewidth{0pt}
\tablecaption{X-ray Properties and Redshifts of the 850~$\mu$m ALMA Sample from C18}
\scriptsize
\tablehead{C18 & L17 & R.A. & Decl. & $f_{850\,\mu m}$ & Error & $f_{\rm 0.5-2\,keV}$ & Error & $f_{\rm 2-7\,keV}$ & Error & $z_{spec}$ & $z_{phot} $ \\ 
No. & No. & J2000.0 & J2000.0 & \multicolumn{2}{c}{(mJy)} & 
\multicolumn{2}{c}{(10$^{-17}$~erg~cm$^{-2}$~s$^{-1})$}  &
\multicolumn{2}{c}{(10$^{-17}$~erg~cm$^{-2}$~s$^{-1})$}   \\ 
(1) & (2) & (3) & (4) & (5) & (6) & (7) & (8) & (9) & (10) & (11) & (12)}
\startdata
1 & 162 &   53.030373 &  -27.855804 &   8.93 & 0.21 &   2.73 &  0.58  &   18.1 &   3.78  & \nodata & [ 3.28]\cr
2 & \nodata &   53.047211 &  -27.870001 &   8.83 & 0.26 & \nodata & \nodata  & \nodata & \nodata  & \nodata &  4.45\cr
3 & 276 &   53.063877 &  -27.843779 &   6.61 & 0.16 &   14.9 &  0.35  &   37.3 &   2.08  & \nodata &  3.12\cr
4 & 132 &   53.020374 &  -27.779917 &   6.45 & 0.41 &   5.86 &  0.61  &   3.86 &   3.66  &  2.252 &  1.95\cr
5 & 522 &   53.118790 &  -27.782888 &   6.39 & 0.16 &   2.38 &  0.35  &   3.79 &   1.77  &  2.309 &  2.29\cr
6 & 844 &   53.195126 &  -27.855804 &   5.90 & 0.18 &   4.94 &  0.48  &   12.9 &   3.03  & \nodata & [ 3.13]\cr
7 & 714 &   53.158371 &  -27.733612 &   5.60 & 0.14 &   7.06 &  0.47  & 0.0331 &   2.91  & \nodata &  3.48\cr
8 & 453 &   53.105247 &  -27.875195 &   5.18 & 0.22 &   2.28 &  0.39  &   2.89 &   2.28  & \nodata &  2.69\cr
9 & 666 &   53.148876 &  -27.821167 &   5.09 & 0.12 &   5.30 &  0.33  &   33.6 &   1.74  &  2.576 &  2.55\cr
10 & 353 &   53.082085 &  -27.767279 &   4.90 & 0.29 &   2.93 &  0.31  &  0.283 &   1.72  & \nodata &  2.41\cr
11 & 342 &   53.079376 &  -27.870806 &   4.76 & 0.29 &   3.25 &  0.44  &   2.37 &   2.61  & \nodata & [ 3.33]\cr
12 & 640 &   53.142792 &  -27.827888 &   4.73 & 0.16 &   2.81 &  0.33  &   4.74 &   1.68  & \nodata &  3.76\cr
13 & \nodata &   53.074837 &  -27.875916 &   4.69 & 0.81 & -0.349 &  0.56  &  0.666 &   3.05  & \nodata & [ 3.68]\cr
14 & 391 &   53.092335 &  -27.826834 &   4.64 & 0.17 &   3.18 &  0.31  &   4.25 &   1.77  & \nodata &  2.73\cr
15 & 140 &   53.024292 &  -27.805695 &   3.93 & 0.15 &   2.54 &  0.48  &   4.12 &   2.99  & \nodata &  2.14\cr
16 & 356 &   53.082752 &  -27.866585 &   4.31 & 0.15 &   3.91 &  0.41  &   14.1 &   2.34  & \nodata &  3.37\cr
17 & 657 &   53.146629 &  -27.871029 &   3.80 & 0.18 &   2.86 &  0.38  &   39.5 &   2.25  & \nodata &  3.57\cr
18 & \nodata &   53.092834 &  -27.801332 &   5.21 & 0.32 &  0.645 &  0.33  & -0.780 &   1.70  &  3.847 &  2.96\cr
19 & 472 &   53.108795 &  -27.869028 &   3.62 & 0.17 &  -1.51 &  0.36  &   9.12 &   2.09  & \nodata & [ 6.69]\cr
20 & 852 &   53.198292 &  -27.747889 &   3.61 & 0.30 &  -3.79 &  0.58  &   11.4 &   3.64  & \nodata &  1.93\cr
21 & \nodata &   53.178333 &  -27.870222 &   3.55 & 0.20 & 0.0829 &  0.51  &   1.44 &   3.08  & \nodata &  3.78\cr
22 & 805 &   53.183460 &  -27.776638 &   3.38 & 0.32 &   54.0 &  0.36  &   53.1 &   2.13  &  2.698 &  2.86\cr
23 & 710 &   53.157207 &  -27.833500 &   3.32 & 0.29 &   2.30 &  0.32  &   6.26 &   1.68  & \nodata &  1.58\cr
24 & \nodata &   53.102791 &  -27.892860 &   3.25 & 0.14 &  0.842 &  0.53  &   4.94 &   3.37  & \nodata &  1.96\cr
25 & \nodata &   53.181377 &  -27.777557 &   3.18 & 0.23 & 0.0756 &  0.35  &   5.40 &   2.00  &  2.794 &  2.92\cr
26 & 299 &   53.070251 &  -27.845612 &   3.15 & 0.25 &   2.26 &  0.34  &   9.20 &   1.98  & \nodata &  3.78\cr
27 & \nodata &   53.014584 &  -27.844389 &   3.05 & 0.19 & 0.0929 &  0.65  &   4.74 &   4.24  & \nodata & [ 4.60]\cr
28 & 622 &   53.139290 &  -27.890722 &   2.89 & 0.37 &   1.38 &  0.50  &   1.25 &   3.20  & \nodata & [ 3.59]\cr
29 & \nodata &   53.137127 &  -27.761389 &   2.82 & 0.28 &   1.23 &  0.31  &   3.50 &   1.69  & \nodata & \nodata\cr
30 & \nodata &   53.071709 &  -27.843693 &   2.78 & 0.15 &  0.703 &  0.33  &   1.47 &   1.89  & \nodata &  1.86\cr
31 & \nodata &   53.077377 &  -27.859612 &   2.54 & 0.43 &   1.36 &  0.40  &   8.80 &   2.21  & \nodata &  1.95\cr
32 & \nodata &   53.049751 &  -27.770971 &   2.56 & 0.16 & -0.187 &  0.45  &  -1.83 &   2.52  & \nodata &  2.75\cr
33 & 313 &   53.072708 &  -27.834278 &   2.49 & 0.23 &   3.29 &  0.33  &  -1.90 &   1.72  & \nodata &  1.58\cr
34 & 386 &   53.090752 &  -27.782473 &   2.47 & 0.21 &   1.00 &  0.33  &   34.9 &   1.65  & \nodata &  1.95\cr
35 & 389 &   53.091747 &  -27.712166 &   2.47 & 0.13 &   2.97 &  0.65  &  -4.38 &   4.34  &  1.612 &  1.71\cr
36 & \nodata &   53.086586 &  -27.810249 &   2.41 & 0.25 &  0.913 &  0.32  &   3.52 &   1.67  & \nodata &  2.37\cr
37 & \nodata &   53.146378 &  -27.888807 &   2.35 & 0.28 &   5.71 &  0.50  &   46.6 &   3.15  & \nodata &  2.96\cr
38 & 393 &   53.092335 &  -27.803223 &   2.50 & 0.10 &   10.5 &  0.33  &   68.5 &   1.70  & \nodata &  2.31\cr
39 & \nodata &   53.124332 &  -27.882696 &   2.26 & 0.18 &   1.61 &  0.45  &   2.37 &   2.61  & \nodata & 0.766\cr
40 & 587 &   53.131123 &  -27.773195 &   2.26 & 0.17 &   5.06 &  0.34  &   21.2 &   1.72  &  2.224 &  2.22\cr
41 & \nodata &   53.172832 &  -27.858860 &   2.25 & 0.18 &   1.12 &  0.39  &  0.156 &   2.39  & \nodata & \nodata\cr
42 & 387 &   53.091629 &  -27.853390 &   2.25 & 0.18 &   3.30 &  0.34  &   53.5 &   1.87  & \nodata &  2.34\cr
43 & \nodata &   53.068874 &  -27.879723 &   2.23 & 0.41 &  0.852 &  0.54  &   2.78 &   3.25  & \nodata & 0.671\cr
44 & \nodata &   53.087166 &  -27.840195 &   2.21 & 0.12 &   1.80 &  0.32  &  0.869 &   1.69  & \nodata & [ 5.33]\cr
45 & 195 &   53.041084 &  -27.837721 &   2.43 & 0.21 &   8.58 &  0.41  &   35.2 &   2.57  & \nodata & [ 3.09]\cr
46 & 449 &   53.104912 &  -27.705305 &   2.29 & 0.11 &   250. &  0.72  &   273. &   4.68  &  1.613 &  1.69\cr
47 & 739 &   53.163540 &  -27.890556 &   2.05 & 0.15 &   2.24 &  0.57  &   31.5 &   3.72  & \nodata &  2.19\cr
48 & 718 &   53.160664 &  -27.776251 &   2.04 & 0.36 &   3.08 &  0.34  &   1.34 &   1.73  &  2.543 &  2.58\cr
49 & 234 &   53.053669 &  -27.869278 &   1.98 & 0.23 &   6.74 &  0.51  &   1.38 &   3.10  & \nodata &  1.87\cr
50 & \nodata &   53.089542 &  -27.711666 &   1.97 & 0.45 &   1.16 &  0.66  &  -3.66 &   4.38  & \nodata &  1.69\cr
51 & \nodata &   53.067833 &  -27.728889 &   1.94 & 0.22 &  0.793 &  0.55  &   6.95 &   3.54  & \nodata &  2.32\cr
52 & \nodata &   53.064793 &  -27.862638 &   1.88 & 0.24 &   1.48 &  0.42  &  0.481 &   2.42  & \nodata & [ 3.63]\cr
53 & 854 &   53.198875 &  -27.843945 &   1.86 & 0.32 &   5.19 &  0.45  &   124. &   2.85  & \nodata &  1.56\cr
54 & 802 &   53.181995 &  -27.814196 &   1.82 & 0.30 &  -1.58 &  0.36  &   7.94 &   1.93  & \nodata & [ 1.85]\cr
55 & \nodata &   53.048378 &  -27.770306 &   1.79 & 0.15 &  0.720 &  0.48  &   1.41 &   2.58  & \nodata & [ 2.76]\cr
56 & 458 &   53.107044 &  -27.718334 &   1.61 & 0.25 &   116. &  0.54  &   391. &   3.55  &  2.299 & [ 2.93]\cr
57 & \nodata &   53.033127 &  -27.816778 &   1.72 & 0.26 &  0.738 &  0.41  &  0.879 &   2.62  & \nodata &  3.08\cr
58 & \nodata &   53.183666 &  -27.836500 &   1.72 & 0.31 & -0.191 &  0.37  &  -1.81 &   2.16  & \nodata & \nodata\cr
59 & 401 &   53.094044 &  -27.804195 &   1.84 & 0.13 &   2.04 &  0.35  &   46.2 &   1.78  &  2.325 &  1.24\cr
60 & \nodata &   53.124584 &  -27.893305 &   1.61 & 0.25 &  0.363 &  0.54  & -0.257 &   3.27  & \nodata &  2.53\cr
61 & \nodata &   53.132751 &  -27.720278 &   1.61 & 0.25 &   1.89 &  0.53  &   2.28 &   3.29  & \nodata & \nodata\cr
62 & \nodata &   53.080669 &  -27.720861 &   1.59 & 0.17 &   2.67 &  0.57  &  -1.06 &   3.61  & \nodata &  2.94\cr
63 & \nodata &   53.120041 &  -27.808277 &   1.57 & 0.26 &   1.41 &  0.29  &   1.08 &   1.64  & \nodata &  1.83\cr
64 & \nodata &   53.117085 &  -27.874918 &   1.53 & 0.31 &   1.54 &  0.38  &   3.76 &   2.28  & \nodata &  3.26\cr
65 & 588 &   53.131458 &  -27.841389 &   1.46 & 0.14 &   1.09 &  0.29  &   13.6 &   1.66  & \nodata &  1.58\cr
66 & 203 &   53.044708 &  -27.802027 &   1.44 & 0.26 &   1.76 &  0.37  &   6.95 &   2.13  & 0.653 & 0.680\cr
67 & 310 &   53.072002 &  -27.819000 &   1.36 & 0.19 &   1.41 &  0.32  &   5.30 &   1.69  & \nodata &  1.69\cr
68 & \nodata &   53.120461 &  -27.742083 &   1.35 & 0.24 & .00277 &  0.41  &   2.69 &   2.21  & \nodata & \nodata\cr
69 & \nodata &   53.113125 &  -27.886639 &   1.25 & 0.27 &  0.477 &  0.46  &   1.84 &   2.84  & \nodata &  2.55\cr
70 & \nodata &   53.141251 &  -27.872860 &   1.18 & 0.25 & 0.0897 &  0.39  &   5.38 &   2.29  & \nodata &  3.14\cr
71 & 245 &   53.056873 &  -27.798389 &   1.16 & 0.30 &   1.81 &  0.34  &   2.48 &   1.85  & \nodata &  1.71\cr
72 & 527 &   53.119957 &  -27.743137 &   1.11 & 0.29 &   3.46 &  0.38  & -0.155 &   2.11  & \nodata & \nodata\cr
73 & \nodata &   53.142872 &  -27.874084 &   1.07 & 0.17 &   1.10 &  0.41  &   2.38 &   2.37  & \nodata &  2.19\cr
74 & 395 &   53.093666 &  -27.826445 &  0.93 & 0.23 &   2.30 &  0.32  &   2.56 &   1.76  & 0.732 & 0.766\cr
75 & \nodata &   53.074837 &  -27.787111 &  0.84 & 0.14 & \nodata & \nodata  & \nodata & \nodata  & \nodata & \nodata\cr
\enddata
\tablecomments{The columns are (1) ALMA source number from C18, 
(2) X-ray source number from L17 when match exists, (3) and (4) ALMA R.A. and Decl.,
(5) and (6) ALMA 850~$\mu$m flux and error from C18, 
(7) and (8) observed 0.5--2~keV flux and error, (9) and (10) observed 2--7~keV flux and error,
(11) specz, and (12) photz from H14, or, if there is no photz in H14, 
then from S16. We provide a FIR-based redshift estimate
from C18 in brackets if there is no specz or high-quality photz available.
}
\label{tab1}
\end{deluxetable}
\onecolumngrid 


\begin{references}

\reference{alex03b}
Alexander, D. M., Bauer, F. E., Brandt, W. N., et al.\ 2003, \aj, 126, 539 

\reference{ashby13}
Ashby, M. L. N., Willner, S. P., Fazio, G. G., et al.\ 2013, \apj, 769, 80

\reference{balestra10}
Balestra, I., Mainieri, V., Popesso, P., et al.\ 2010, A\&A, 512, 12

\reference{barger15}
Barger, A. J., Cowie, L. L., Owen, F., et al.\ 2015, \apj, 801, 87

\reference{bcs01}
Barger, A. J., Cowie, L. L., Steffen, A. T., et al.\ 2001, \apj, 560, L23

\reference{blain93}
Blain, A. W, \& Longair, M. S.\ 1993, \mnras, 264, 509

\reference{casey11}
Casey, C. M., Chapman, S. C., Smail, I., et al.\ 2011, \mnras, 411, 2739

\reference{cooper12}
Cooper, M. C., Yan, R., Dickinson, M., et al.\ 2012, \mnras, 425, 2116

\reference{cowie20}
Cowie, L. L., Barger, A. J., Bauer, F. E., \& Gonz{\'a}lez-L{\'o}pez, J.\ 2020,
\apj, in press

\reference{cowie12}
Cowie, L. L., Barger, A. J., \& Hasinger, G.\ 2012, \apj, 748, 50

\reference{cowie17}
Cowie, L. L., Barger, A. J., Hsu, L.-Y., et al.\ 2017, \apj, 837, 139 (C17)

\reference{cowie16}
Cowie, L. L., Barger, A. J., \& Songaila, A.\ 2016, \apj, 817, 57

\reference{cowie18}
Cowie, L. L., Gonz{\'a}lez-L{\'o}pez, J., Barger, A. J., et al.\ 2018, \apj, 865, 106 (C18)

\reference{cowie96}
Cowie, L. L., Songaila, A., Hu, E. M., \& Cohen, J. G.\ 1996, \aj, 112, 839

\reference{dahlen13}
Dahlen, T., Mobasher, B., Faber, S. M., et al.\ 2013, \apj, 775, 93

\reference{dunlop17}
Dunlop, J. S., McLure, R. J., Biggs, A. D., et al.\ 2017, \mnras, 466, 861

\reference{faber03}
Faber, S. M., Phillips, A. C., Kibrick, R. I., et al.\ 2003, SPIE, 4841, 1657

\reference{fixsen98}
Fixsen, D. J., Dwek, E., Mather, J. C., Bennett, C. L., \& Shafer, R. A.\ 1998,
\apj, 508, 123

\reference{franco18}
Franco, M., Elbaz, D., B{\'e}thermin, M., et al.\ 2018, A\&A, 620, A152

\reference{harrison12}
Harrison, C. M., Alexander, D. M., Mullaney, J. R., et al.\ 2012, \apj, 760, L15

\reference{hatzim10}
Hatziminaoglou, E., Omont, A., Stevens, J. A., et al.\ 2010, A\&A, 518, L33

\reference{hodge13}
Hodge, J. A., Karim, A., Smail, I., et al.\ 2013, \apj, 768, 91

\reference{hopkins06}
Hopkins, P. F., Hernquist, L.,?Cox, T. J., et al.\ 2006, \apjs, 163, 1

\reference{hsieh12}
Hsieh, B.-C., Wang, W.-H., Hsieh, C.-C., et al.\ 2012, \apjs, 203, 23

\reference{hsu14}
Hsu, L.-T., Salvato, M., Nandra, K., et al.\ 2014, \apj, 796, 60 (H14)

\reference{inami17}
Inami, H., Bacon, R., Brinchmann, J., et al.\ 2017, A\&A, 608, A2

\reference{kriek15}
Kriek, M., Shapley, A. E., Reddy, N. A., et al.\ 2015, \apjs, 218, 15

\reference{kroupa01}
Kroupa, P.\ 2001, \mnras, 322, 231

\reference{kurk13}
Kurk, J., Cimatti, A., Daddi, E., et al.\ 2013, A\&A, 549, 63

\reference{labbe15}
Labb{\'e}, I., Oesch, P. A., Illingworth, G. D., et al.\ 2015, \apjs, 221, 23 

\reference{lehmer12}
Lehmer, B. D., Xue, Y. Q., Brandt, W. N., et al.\ 2012, \apj, 752, 46

\reference{lefev05}
Le F{\`e}vre, O, Vettolani, G., Garrili, B., et al.\ 2005, A\&A , 439, 845

\reference{luo17}
Luo, B., Brandt, W. N., Xue, Y. Q., et al.\ 2017, \apjs, 228, 2 (L17)

\reference{lutz10}
Lutz, D., Mainieri, V., Rafferty, D., et al.\ 2010, \apj, 712, 1287

\reference{mclean12}
McLean, I. S., Steidel, C. C., Epps, H. W., et al.\ 2012, SPIE, 8446, 84460J

\reference{mclure18}
McLure, R. J., Pentericci, L., Cimatti, A., et al.\ 2018, \mnras, 479, 25

\reference{mignoli05}
Mignoli, M., Cimatti, Z., Zamorani, G., et al.\ 2005, A\&A, 437, 883

\reference{mineo12}
Mineo, S., Gilfanov, M, \& Sunyaev, R.\ 2012, \mnras, 419, 2095

\reference{morris15}
Morris, A. M., Kocevski, D. D., Trump, J. R., et al.\ 2015, \aj, 149, 178

\reference{mull15}
Mullaney, J. R., Alexander, D. M., Aird, J., et al.\ 2015, \mnras, 453, L83

\reference{nguyen10}
Nguyen, H. T., Schulz, B., Levenson, L., et al.\ 2010, A\&A, 518, L5

\reference{oke95}
Oke, J. B., Cohen, J. G., Carr, M., et al.\ 1995, PASP, 107, 375

\reference{page12}
Page, M. J., Symeonidis, M., Vieira, J. D., et al.\ 2012, Natur, 485, 213

\reference{popesso09}
Popesso, P., Dickinson, M., Nonino, M., et al.\ 2009, A\&A, 494, 443

\reference{rafferty11}
Rafferty, D. A., Brandt, W. N., Alexander, D. M., et al.\ 2011, \apj, 742, 3

\reference{rama19}
Ramasawmy, J., Stevens, J., Martin, G., Geach, J. E.\ 2019, \mnras, 486, 4320

\reference{rosario12}
Rosario, D. J., Santini, P., Lutz, D., et al.\ 2012, \aa, 545, A45

\reference{sanders88}
Sanders, D. B., Soifer, B. T., Elias, J. H., et al.\ 1988, \apj, 325, 74

\reference{santini15}
Santini, P., Ferguson, H. C., Fontana, A., et al.\ 2015, \apj, 801, 97

\reference{santini09}
Santini, P., Fontana, A., Grazian, A., et al.\ 2009, A\&A, 504, 751

\reference{saz17}
Sazonov, S., \& Khabibullin, I.\ 2017, \mnras, 468, 2249

\reference{scholtz18}
Scholtz, J., Alexander, D. M., Harrison, C. M., et al.\ 2018, \mnras, 475, 1288

\reference{schreiber17}
Schreiber, C. Pannella, M., Leiton, R., et al.\ 2017, A\&A, 599, A134

\reference{shao10}
Shao, L, Lutz, D., Nordon,R., et al.\ 2010, A\&A, 518, L26

\reference{silverman10}
Silverman, J. D., Mainieri, V., Salvato, M., et al.\ 2010, \apjs, 191, 124

\reference{skelton14}
Skelton, R. E., Whitaker, K. E., Momcheva, I. G., et al.\ 2014, \apjs, 214, 24

\reference{straatman16}
Straatman, C. M. S., Spitler, L. R., Quadri, R. F., et al.\ 2016, \apj, 830, 51 (S16)

\reference{stanley15}
Stanley, F., Harrison, C. M., Alexander, D. M., et al.\ 2015, \mnras, 453, 591

\reference{stanley18}
Stanley, F., Harrison, C. M., Alexander, D. M., et al.\ 2018, \mnras, 478, 3721

\reference{szokoly04}
Szokoly, G. P., Bergeron, J., Hasinger, G., et al.\ 2004, \apjs, 155, 271

\reference{ueda18}
Ueda, Y., Hatsukade, B., Kohno, K., et al.\ 2018, \apj, 853, 24

\reference{urrutia19}
Urrutia, T., Wisotzki, L., Kerutt, J., et al.\ 2019, A\&A, 624, A141

\reference{vanzella08}
Vanzella, E., Cristiani, S., Dickinson, M., et al.\ 2008, A\&A, 478, 83

\reference{xue11}
Xue, Y. Q., Luo, B., Brandt, W. N., et al.\ 2011, \apjs, 195, 10

\reference{zavala17}
Zavala, J. A., Aretxaga, I., Geach, J. E., et al.\ 2017, \mnras, 464, 3369

\end{references}
\end{document}